# Terrestrial planet and asteroid belt formation by Jupiter–Saturn chaotic excitation

*Patryk Sofia Lykawka[1] & Takashi Ito[2,3]

[1] Kindai University, Shinkamikosaka 228-3, Higashiosaka, Osaka, 577-0813, Japan; patryksan@gmail.com

[2] Center for Computational Astrophysics, National Astronomical Observatory of Japan, Osawa 2-21-1, Mitaka, Tokyo 181-8588, Japan

[3] Planetary Exploration Research Center, Chiba Institute of Technology, 2-17-1 Tsudanuma, Narashino, 275-0016, Chiba, Japan

## Abstract

The terrestrial planets formed by accretion of asteroid-like objects within the inner solar system's protoplanetary disk. Previous works have found that forming a small-mass Mars requires the disk to contain little mass beyond ~1.5 au (i.e., the disk mass was concentrated within this boundary). The asteroid belt also holds crucial information about the origin of such a narrow disk. Several scenarios may produce a narrow disk. However, simultaneously replicating the four terrestrial planets and the inner solar system properties remains elusive. Here, we found that chaotic excitation of disk objects generated by a near-resonant configuration of Jupiter–Saturn can create a narrow disk, allowing the formation of the terrestrial planets and the asteroid belt. Our simulations showed that this mechanism could typically deplete a massive disk beyond ~1.5 au on a 5–10 Myr timescale. The resulting terrestrial systems reproduced the current orbits and masses of Venus, Earth and Mars. Adding an inner region disk component within ~0.8–0.9 au allowed several terrestrial systems to simultaneously form analogues of the four terrestrial planets. Our terrestrial systems also frequently satisfied additional constraints: Moon-forming giant impacts occurring after a median ~30–55 Myr, late impactors represented by disk objects formed within 2 au, and effective water delivery during the first 10–20 Myr of Earth's formation. Finally, our model asteroid belt explained the asteroid belt's orbital structure, small mass and taxonomy (S-, C- and D/P-types).

**Introduction**

Early terrestrial-planet formation models postulated that the terrestrial planets formed in a protoplanetary disk that extended until the disk outer edge at ~4-5 au (ref.[1,2]). However, it was later found that these models could not explain the small mass of Mars[1,2,3,4,5]. Since this constraint is essential to explaining the inner solar system, we focus only on models capable of explaining Mars' mass in this work. A common view in recent studies is that ~2 Earth masses (ME) of mass concentrated in a narrow disk at 0.7–1.0 au to form Venus and Earth[6]. In this paradigm, nearly formed Mercury and Mars were scattered out from this region and remained of low mass due to a lack of disk mass within 0.7 au and beyond 1 au (ref.[7,8,9,10]). Other models suggest that both planets formed across a similarly narrow disk in tandem with the Venus–Earth pair by the end of disk gas dispersal[11]. A narrow disk could have resulted from disk-gas-driven convergence of small bodies during the solar system's first Myr, resulting in highly mass concentrated rings near the Venus–Earth region[12,13,14]. Alternatively, disk-gas-driven inward-then-outward Jupiter migration (Grand Tack) or giant-planet instability (henceforth 'instability') could also have created narrow disks by dynamical truncation of the disk beyond ~1.5 au (ref.[8,9]). Another key question is how the origin of this narrow disk (or the terrestrial planet system) is related to asteroid belt formation. The asteroid belt consists of asteroids concentrated at semimajor axes $a = 2$–$3.25$ au possessing a wide range of eccentricities ($e < 0.4$) and inclinations ($i < 35°$). Furthermore, the total mass of the asteroid belt is only $5 \times 10^{-4}$ ME (ref.[15]). It is unclear whether the primordial asteroid belt was dynamically depleted (e.g., by Jupiter)[8,16,17] or was not massive originally[12,13,18]. The asteroid belt holds essential clues regarding the nature of the protoplanetary disk that formed the terrestrial planets.

In addition to the orbits and masses of the four terrestrial planets, other essential constraints in the inner solar system include the planets' formation timescales and accretion history (e.g., giant impacts), Moon formation (e.g., timing) and mass accreted by Earth after that, nature of the planets' late impactors, origin and accretion evolution of water on all the four terrestrial planets, among others. Furthermore, the absence of planets, the orbital architecture (including the peculiar concentration of asteroids with $i < 20$ deg), compositional taxonomy, and the low mass in the asteroid belt represent additional fundamental constraints (Supplementary Information 1). While the Grand Tack[8] and the early instability[6,9] models investigated several constraints related to the terrestrial planets and the asteroid belt, other models[7,10,11,13] addressed these issues in much less depth (e.g., focusing only on a few chosen constraints from the list above). As discussed below, several models in the literature also neglected Mercury formation (e.g., the Grand Tack). Furthermore, forming the asteroid belt in tandem with the terrestrial planets remains poorly understood. Therefore, there has been insufficient discussion about explaining *simultaneously* the orbits, masses and other constraints of the four terrestrial planets and the asteroid belt in the literature[6,16,19,20,21].

Noteworthy, only a few inner solar system models have tackled the formation of terrestrial planets and the asteroid belt in a single evolutionary fashion (e.g., the Grand Tack and early instability models). Despite these models' new insights and successes, we discuss several issues related to them that are absent or possibly less relevant in our model. We also argue that the 0.7–1.0 au narrow disk (henceforth 'canonical annulus') model is not an adequate baseline for terrestrial planet formation. Other models face important challenges and lack discussions regarding the inner solar system constraints, so they are not discussed below. See Supplementary Information 2 for details. In this work, we considered the simultaneous formation of terrestrial planets and the asteroid belt. Namely, we addressed all the inner solar system constraints described above with unprecedented detail, thus making our model substantially comprehensive.

We performed N-body simulations of the early solar system consisting of the giant planets and an extended massive protoplanetary disk after gas dispersal. In our 650 main simulations, the pre-instability Jupiter–Saturn pair experienced their mutual near 2:1 mean motion resonance (MMR) on moderately eccentric orbits. The Methods section demonstrates that this orbital configuration was plausible and probably arose naturally before the instability. In particular, the presence of Mars-Earth-mass bodies or additional planets in the outer solar system probably played a role in originating this orbital configuration. Noting that the instability probably occurred in ~10 Myr timescales after the solar system's gas dispersal[22,23,24] (see also Methods), for simplicity, we assumed that this near-resonant stage operated in similar timescales before the instability in our simulations. In the next stage (post-instability), we took the orbital state of all disk objects at the end of the previous stage and placed Jupiter and Saturn on their near-current orbits. We justify this simplification of the instability in Methods. We then followed the evolution of these terrestrial systems until 400 Myr. In these simulations, the protoplanetary disk consisted of a small number of embryos (Moon-Mars-mass objects) and several planetesimals (small asteroid-like objects) that concentrated at smaller distances and up to 3.5 au in the disk, respectively. Consistent with predictions of embryo/planetesimal formation models, implications from our previous results, and some fundamental constraints about the terrestrial planets, we tested several variations in disk properties. This procedure resulted in our distinct disk models, as illustrated in Figure S1 and summarised in Table S1. The disks comprised a core region surrounded by a less massive inner region and an extended outer region. In all disks modelled, the embryos and planetesimals started on nearly circular and coplanar orbits. The disk component beyond 2 au consisted of planetesimals and represented the primordial asteroid belt (local asteroids), which was thousands of times more massive than the current. We also investigated the existence of spectral classes inspired by asteroid taxonomy (S, C and D/P types) and distinct water mass fractions for our disk objects (Methods and Table S2). Finally, we used a rigorous classification algorithm to properly identify analogues of the terrestrial planets (Methods and Supplementary Information 2.1).

Furthermore, we used additional long-term simulations to investigate the formation of the asteroid belt. Specifically, we built a representative asteroid belt consisting of local and captured asteroids. The local asteroids were obtained in systems containing good representatives of the terrestrial planets from the main simulations described above. The captured asteroids were obtained from simulations of trans-Jovian objects captured in the asteroid belt during the instability/migration of the giant planets. Finally, we took the orbital states of local and captured asteroids at $t \sim 100$ Myr and evolved them until 4 Gyr. We also tested the influence of the instability and post-instability residual migration on our results based on auxiliary simulations. Figure 1 summarises the timeline of the main events envisioned in our scenario.

Consult Methods and Supplementary Information 3 for more details about our simulations and their initial conditions.

## Results and Discussions

Truncating the protoplanetary disk before the instability

First, when placing the Jupiter–Saturn pair on a near 2:1 MMR configuration, Saturn's natural frequencies associated with the longitude of perihelion and the longitude of ascending node wander chaotically over a wide range of values (Figure S2 and Supplementary Information 4). Thus, several asteroids are affected by the associated nu6 and nu16 secular resonances resulting in widespread orbital excitation in the asteroid belt[25]. Here, we found that this Jupiter–Saturn chaotic excitation (JSCE) mechanism offers a new route to generate a narrow disk and has far-reaching implications for the inner solar system. Considering that the strength of the nu6 resonance is sensitively proportional to planetary eccentricities and because the orbits of our Jupiter–Saturn pairs were more eccentric than that assumed in past work[25], a stronger nu6 resonance operated in the disk[4]. As a result, JSCE strongly depleted the system's disk beyond ~1–1.5 au (up to ~3.5 au), as most asteroids acquired large eccentricities leading to gravitational scattering by the giant planets and collisions with the Sun (Figures S3 and S4). These results demonstrate that moderately eccentric Jupiter and Saturn experiencing JSCE could both excite and deplete the primordial asteroid belt. Consequently, violent instabilities (e.g., the early instability model) or complex Grand-Tack-like scenarios are not required to truncate a massive extended disk. It is also unnecessary to assume that the disk was initially narrow[7,18] (i.e., with "empty" outer regions beyond ~1 au), which assumes that planetesimal formation occurred only under special conditions at a specific desirable and narrow region of the disk[11,12,13,14,26,27]. Instead, the only requirement in the JSCE scenario is that moderately eccentric Jupiter and Saturn experience their 2:1 MMR prior to the instability. Thus, our scenario probably requires less fine-tuning to operate. Another implication is that the instability occurred when the disk was already strongly perturbed beyond ~1 au.

Terrestrial planet analogue systems

Our terrestrial planet classification algorithm carefully identified terrestrial planets/systems. In total, our main simulations revealed 221 terrestrial systems containing at least Venus, Earth and one additional planet analogue in each system. Among these, we identified a significant number of systems containing planet analogues of Mercury, Venus, Earth and Mars in the same system (47) (henceforth '4-P system'), which closely resembled the four terrestrial planets in terms of orbits and masses (Figures 2 and 3). Furthermore, these systems were excellent representatives of our own because they satisfied many additional vital constraints of the inner solar system (Introduction). Remarkably, some systems could satisfy several of these constraints *simultaneously*. Overall, our Venus and Earth analogues acquired dynamically cold orbits, while our Mercury and Mars analogues acquired hotter orbits. Indeed, their median *a-e-i* orbital elements were very similar to those of the real planets. The orbit-mass distribution of these analogues also reproduced the dichotomy of two massive planets (Venus and Earth) surrounded by two much less massive planets (Mercury and Mars) on relatively distant orbits. Other terrestrial system properties are summarised in Supplementary Information 5.1 and Tables S3, S4 and S5. Although our Mercury–Venus pairs formed on average closer to each other than in reality, our Venus–Earth and Earth–Mars pairs matched observations quite well. Regardless, these mutual orbital separations were better than those obtained for narrow disks based on the canonical annulus model (Supplementary Information 2.2 and 5.2). Finally, improving our pioneer research on 4-P systems[19], we obtained a significant number of 4-P systems analogous to the terrestrial planets and statistically compared them to many inner solar system constraints.

Formation of Mercury, Venus, Earth and Mars

Although the formation of Mercury is an outstanding problem often neglected in the literature (Supplementary Information 2.3), our model produced a significant number of Mercury analogues (122) (Table S4) belonging to systems analogous to the inner solar system, of which 47 were found in 4-P systems. In addition to increasing the chances of Mercury formation with high efficiency (84/250 = 34%), the inclusion of extended inner regions at 0.3–0.85 au (combined disk Ix) also produced more 4-P systems with 13% efficiency (33/250) compared to 3.5% (14/400) for other disks combined (disk A–E) (Methods). These results significantly improved the probabilities of successfully forming Mercury and 4-P systems compared to past work (Supplementary Information 2.2 and 2.3). Mercury analogues acquired medians $a = 0.44$ (more consistent with $a = 0.39$ au for Mercury) and 0.49 au, $e = 0.07$ and 0.11, $i = 3.5°$ and 4.7° and $m = 0.16$ and 0.14 ME for disks Ix and A–E, respectively. In particular, disks containing 0.20–0.25 ME inner regions yielded analogue masses within only a factor of 2 of that of Mercury (disks B, C, Ia, Ic in Table S4). Thus, slightly less massive inner regions could explain Mercury's current mass. Mercury's current hot orbit ($e = 0.21$; $i = 6.8°$) can be explained by post-instability residual migration of the giant planets (Supplementary Information 3.3) or Gyr-long-term chaotic dynamics[28]. Overall, disk Ix produced our best Mercury analogues. These analogues started within the inner region with typically 10% of their final masses. Later, they accreted the remaining 35% and 55% from objects in the inner and distant regions (> 0.85 au), respectively. In particular, the contribution of the inner region to their final masses was ~3 and ~5.5 times higher compared to that for analogues of Venus and Earth/Mars. Also, 10% of the analogues' final masses originated from objects located beyond 1.5 au (but <1% beyond 2 au), implying that these objects probably sourced water and other volatile materials to Mercury, in agreement with recent measurements[29,30]. This result supports the hypothesis that the disk was not "empty" beyond 1.5 au. Such a significant contribution of local and mixing of distant objects could also offer new insights into the origin of Mercury's peculiar physical properties. JSCE did not shape the inner region but affected Mercury's formation (e.g., delivery of water, accretion history of distant objects, etc.). In short, forming Mercury analogues is an essential *additional result* that makes our scenario more comprehensive (see also Supplementary Information 5.2). We conclude that the formation of Mercury probably required a low-mass inner region and a disk outer region extending beyond 1 au in the disk.

In general, our analogue systems contained Venus–Earth pairs with orbits and masses in agreement with the real planets (Table S4). The Venus and Earth analogues acquired medians of $a \sim$ 0.65–0.67 au (Venus), $a \sim$ 1.00–1.07 au (Earth) and $e \sim$ 0.03–0.04, $i \sim 2°$ and $m \sim$ 0.82–1.03 ME for both planets. Concerning the mutual distance of Venus and Earth ($a_E$-$a_V$ = dVE = 0.28 au), a Venus–Earth pair was considered successful if their mutual distance fell within the range 0.67–1.33 × dVE. Approximately 40–50% and 70% of the Venus–Earth pairs acquired successful mutual distances in disks considering initial mass concentrations within 0.8–1.0 au (disks B and C) and 0.85–0.95 au (disk Ix), respectively. Thus, reproducing the Venus–Earth distance likely requires more than 1 ME of mass concentrated in a thin annulus within the disk's core region. This peculiar feature might also explain the small Venus and Earth analogues' eccentricities (Table S4). These results roughly match the Venus–Earth pair's cold orbits, masses and orbital separation. This accomplishment is significant because it is still challenging to correctly reproduce the Venus–Earth's orbit-mass distribution (ref.[31] and Supplementary Information 2). Finally, our results are consistent with the terrestrial planets' late accretion and planetary bulk compositions of both Earth and Mars. Further details are discussed in Supplementary Information 5.3, 5.4 and 5.5.

Our model was highly successful regarding Mars formation, as 146 Mars analogues acquired medians of $a \sim$ 1.55–1.58 au, $e \sim$ 0.07–0.09, $i \sim 4°–6°$ and $m \sim$ 0.13–0.20 ME in our analogue systems (Table S4). JSCE's mass depletion and orbital stirring beyond 1–1.5 au that arose on ~5–10 Myr timescales allowed many Mars analogues to acquire small masses and non-cold orbits in similar timescales. Both the Grand Tack and the early instability models require extreme perturbations on short timescales and precise timings to provide the necessary disk depletion to explain the small mass of Mars. In contrast, no such perturbations are required in our scenario to form Mars. The existence of a massive outer region also caused our Mars analogues to form farther from their Earth counterparts, in agreement with observations. Models in which an outer region beyond ~1 au starts mostly "empty" tend to form Mars too close to Earth (Supplementary Information 2). Briefly, our model can reproduce the moderately excited orbit of Mars, its small mass and its mutual distance from Earth. In contrast, other models typically focus on the small-mass problem only[6,7,8,9,10,11,13,18,32,33,34,35]. Our Mars analogues accreted 80% (90%) of their final masses after a median of 15–18 Myr (20–27 Myr). These timescales are a bit long but marginally consistent with the ~15–23 Myr bulk formation timescale of Mars. The analogues also experienced giant impacts during their accretion histories, which is in line with evidence suggesting that Mars formed in a protracted manner. See also the discussion in Supplementary Information 5.6.

Concerning Moon formation and late veneer mass delivered to Earth, several of our Earth analogues experienced successful late Moon-forming giant impacts (GIs) (Table S5). For a plausible impactor-to-target-mass ratio (ITR) greater than 0.05 (0.02), 55–60% (55–80%) of these analogues experienced GIs after the minimum 25 Myr. Even for the canonical ITR > 0.10, the successful fraction was relatively high: 35–40%. For ITR > 0.05 (0.02), the last GIs occurred after medians of 30–45 (45–55) Myr. Although the Moon-forming timing has large uncertainties (Supplementary Information 1), these results are consistent with a < 60-Myr Moon-forming interval based on isotopic systematics[36] and the 45-Myr timing derived from hafnium-tungsten chronometry[37]. After the last GI, overall 30–40% (40–50%) of our Earth analogues accreted less than 1% of their final masses for ITR > 0.05 (0.02), whilst for ITR > 0.10 the results were 20–25%. It is clear from these results that less massive Moon-forming impactors with a mass of 0.02 ME < $m$ < 0.1 ME can increase the success rates of satisfying both the Moon-formation timing and Earth's late veneer mass constraints. Such conditions are consistent with models of Moon formation. These success rates are also higher than those obtained for the canonical annulus and similar models (Supplementary Information 2.2 and 2.6). As JSCE stirred the orbits of objects located beyond ~1–1.5 au, collisions with Earth became less likely, and Moon-forming GIs occurred later, resulting in fewer remnant objects to later accrete.

Furthermore, after considering 80 water mass fraction (WMF) models that probed different water contents in specific regions of the disk (Table S2), we found that the often assumed WMF models used in the literature[4,9,19] (models 1, 2 and 16 in Table S2) failed to satisfy the water contents of Venus, Earth and Mars simultaneously. Instead, objects with higher WMFs beyond ~1–1.5 au were required to satisfy this constraint (Supplementary Information 5.7). These results imply that this region was relatively massive and wetter than previously thought. Such conditions provided enough water content to the terrestrial planets. The results are also consistent with recent observations of water on enstatite chondrite meteorites[38] and S-type asteroids[39], and models of water-rich objects implanted into the primordial asteroid belt[40]. The amount of water delivered to an Earth analogue was a median of 2–5 times the mass in the oceans. This result lies within the typical ~2–10 oceans estimated for Earth's water mass[38,41]. Therefore, the origin of Earth's bulk water could be explained by collisions with water-rich objects located beyond ~1.5–2 au after being quickly stirred by JSCE. This process allowed Earth's water to be delivered in less than 10–20 Myr of the planet's accretion history. These results agree with isotopic evidence that most of Earth's water was acquired early (before the Moon-forming GI) and not during late accretion[41]. Finally, in agreement with isotopic constraints[42], we found that the Earth analogues acquired 80% (90%) of their final masses within 20 Myr (35 Myr) and the remaining mass in longer timescales, often experiencing giant impacts. In short, the JSCE model can explain the Earth's fast water delivery and protracted formation.

Formation of the asteroid belt

Our scenario successfully satisfied several constraints in the asteroid belt. Before the onset of the instability, a moderately eccentric Jupiter sculpted the primordial asteroid belt beyond ~3.2 au, and JSCE dynamically depleted and stirred the local asteroids, so a tiny fraction of them survived at 2–3.25 au. Our local asteroids were obtained in analogue systems containing three or four representative terrestrial planets from simulations of the standard disk (Figure 3). Later, the instability and migration of the giant planets destabilised many objects in trans-Jovian reservoirs, a small fraction of which contaminated the primordial asteroid belt[24,43]. At the end of giant-planet migration, captured asteroids from these reservoirs survived in the outskirts of the belt[43,44]. After ~4.1 Gyr of evolution, our representative asteroid belt consisted of mixed populations of local and captured asteroids (see Introduction and Methods for details). Next, we compared these results with observations by considering large asteroids not belonging to asteroid families, which represented the asteroid belt orbital structure well. Overall, as evinced by the similarity of the *a-e-i* distributions, the orbital structure of our representative asteroid belt broadly reproduced observations (Figure 4). Our asteroid belt model also featured an unprecedented high resolution compared with similar works[16,17,18,45]. In agreement with ref.[25], the orbital spectra of Saturn when JSCE was active revealed that nu6 resonance was more prominent than nu16 resonance (Figure S2). This behaviour explained the wide range of eccentricities and less-excited inclinations acquired by stable local asteroids. Notably, captured asteroids on stable orbits tend to concentrate below 20° (ref.[43]). Overall, these results may explain the fraction of asteroids with inclinations above the nu6 resonance (f6). Here, after considering the effects of instability, the final stages of giant planet migration, and ~4 Gyr long-term evolution, we obtained f6 = 0.11–0.18, which was smaller than the values obtained previously in the literature. These results are marginally compatible with the observed value of 0.07. Therefore, the JSCE scenario may solve the longstanding problem of reproducing f6. More details and other advantages of our asteroid belt model are found in Supplementary Information 2.5 and 5.8. Overall, we conclude that the good match of the orbital structure of our asteroid belt with the observed one is robust.

JSCE depleted 99.87% of local asteroids in the primordial asteroid belt. This result is better than most previous models predicting values ~90–99% over the age of the solar system[9,17] and comparable to ~99–99.9% found in the instability model for the asteroid belt[16]. However, all depleted systems with > 97% found in the latter work put Jupiter and Saturn on too-excited or mutually distant orbits (orbital period ratio PS/PJ > 2.8). The final masses of our asteroid belt considering both local and captured asteroids varied within $1-2 \times 10^{-3}$ ME, or ~2–4 times the mass in the current asteroid belt, assuming 33–67% of local asteroids represent the latter. This result is compatible with observations because these masses are likely upper limits and could be smaller after considering a less massive primordial asteroid belt or additional depletion mechanisms. Concluding, in addition to the bulk orbital structure, our model can also explain the current small mass of the asteroid belt.

Finally, by considering a primordial asteroid belt consisting of local S- and C-asteroids and captured asteroids akin to C- and D/P-asteroids, we found that our asteroid belt reproduced not only the dichotomy of S- and D/P-type asteroids spread in the inner-middle and outer regions of the belt but also the distribution of C-type asteroids beyond ~2.3 au (ref.[15]). Furthermore, the orbital concentrations of our obtained S-, C- and D/P-asteroids approximately matched the concentrations (peaks) with the heliocentric distance of observed asteroids (Figure 4). Overall, the best results were for an asteroid belt represented by local and captured asteroids in similar proportions (not exceeding a factor of two). The asteroid belt model within the JSCE scenario is the first to successfully explain the orbital distribution of three taxonomic populations represented by S-, C- and D/P-type asteroids.

After the giant planets' instability/migration and Moon formation, the evolution of asteroids in our long-term simulations indicated that roughly 0.0018 ME of mass was added to the nearly formed Earth via asteroid collisions over 4 Gyr. This value is similar to those based on the constraints of highly siderophile elements[46,47]. As JSCE's disk planetesimal depletion was faster beyond ~2 au, long-lived reservoirs remained within this boundary. Indeed, the decaying population of late Earth impactors comprised asteroids concentrated at 1–2 au and $i = 15°–50°$ at the end of the terrestrial-planet formation. Considering that this population originally formed at ~0.8–1.8 au in the disk, these results are consistent with isotopic evidence indicating that most of Earth's late impactors had a dry chondrite nature[48]. Thus, the JSCE scenario can potentially explain Earth's late veneer mass and the nature of late impactors. Finally, our results also imply that these terrestrial-region asteroids (< 2 au) bombarded the entire inner solar system, in agreement with the findings of ref.[46]. Consult Methods and Supplementary Information 5.8 for further details about asteroid belt formation in our scenario.

Summary

The JSCE model could explain a large number of inner solar system constraints. Tentative implications of our scenario are summarised below. First, at the end of gas dispersal, the protoplanetary disk that formed the terrestrial planets was probably massive and comprised three main components: a low-mass (< 0.20–0.25 ME) and extended inner region that was crucial for the formation of Mercury; a massive (~2 ME) core region at ~0.8–1.2 au possessing a mass concentration within a thin layer (> 1 ME within dr < 0.2 au) that allowed the formation of Venus and Earth on mutually close orbits; and a massive adjacent outer region (up to 3.5 au) containing the primordial asteroid belt that was strongly excited/depleted by JSCE over ~5–10 Myr timescales before the instability, which later allowed the formation of Mars. Finally, the current asteroid belt consists of remnant primordial asteroids and captured asteroids from trans-Jovian regions. See also Supplementary Information 6.

The main highlights of our work based on the JSCE scenario are the following.

- A new mechanism to mass depletes a protoplanetary disk is proposed. It is based on the Jupiter–Saturn chaotic excitation generated when both planets evolve on moderately eccentric orbits near their 2:1 MMR before giant planet instability;
- New scenarios to explain a small-mass Mars and a low-mass asteroid belt are proposed as natural consequences of the disk depletion of planetesimals beyond ~1.5 au;
- A new model of terrestrial planet formation is proposed based on the same disk above and the addition of an inner region within ~0.8–0.9 au. After ~100 Myr of post-instability evolution, this model forms terrestrial systems containing the four terrestrial planets with orbits and masses similar to the observed;
- The Earth analogues obtained in the modelled systems often satisfy important constraints: Moon

formation timing, early accretion of water-rich asteroids (Earth's bulk water), and late accretion of dry asteroids;
- A new model of asteroid belt formation is proposed based on the admixing of local and captured asteroids that survived ~4 Gyr of evolution after terrestrial planet formation. Local asteroids are remnant disk objects after JSCE depletion. Captured asteroids contaminated the disk from trans-Jovian reservoirs after the instability. This model broadly explains the asteroid belt's low mass, orbital distribution ($a$-$e$-$i$), and composition taxonomy of three major populations (S-, C-, and D/P-types);
- From the results above, we infer that the inner solar system's bombardment was probably caused by remnant dry asteroids from the tail-end of terrestrial planet formation.

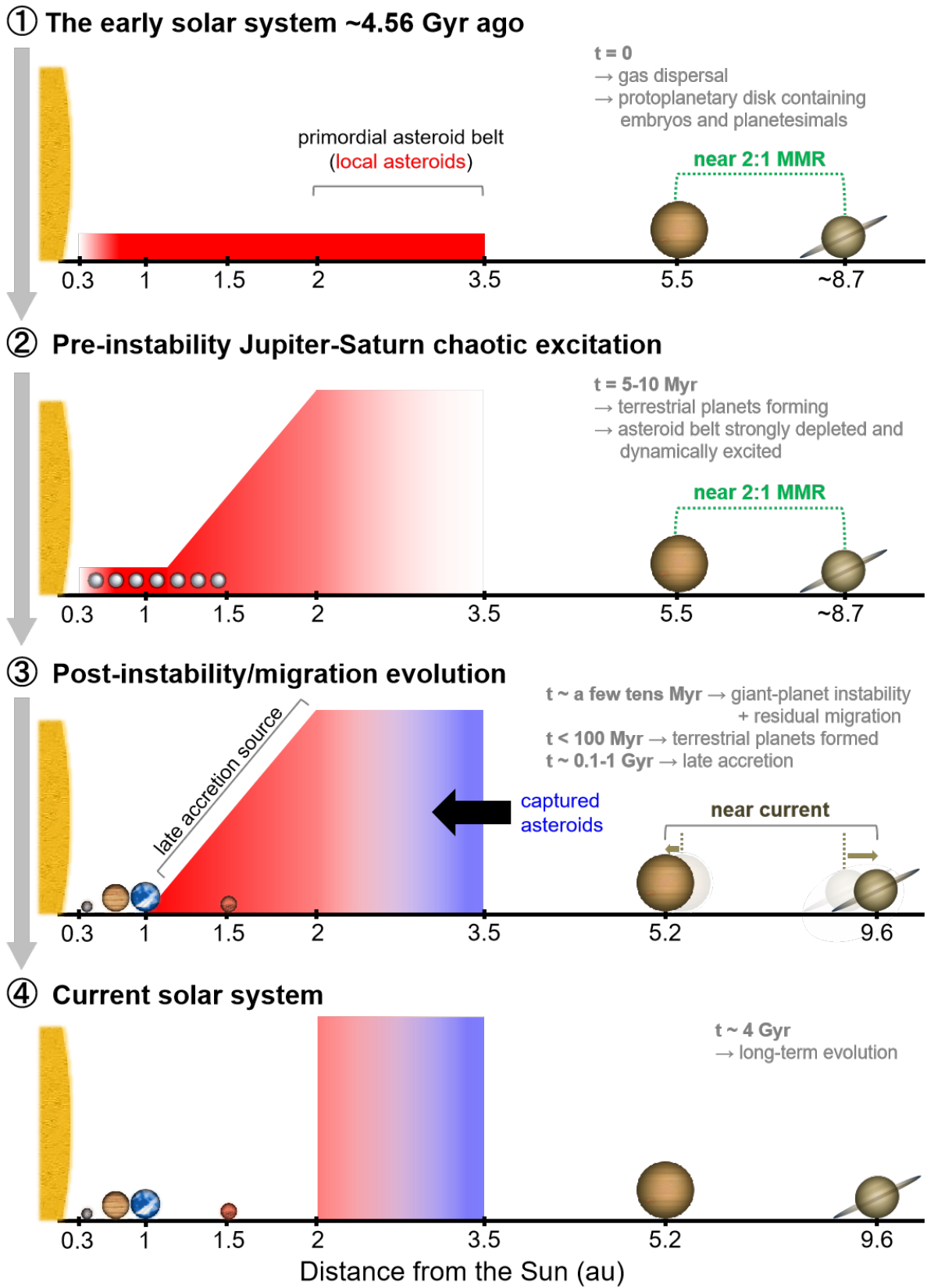

**Figure 1.** Outline of the Jupiter–Saturn chaotic excitation scenario for the formation of the terrestrial planets and the asteroid belt. Four main stages describe the dynamical history of the solar system. The approximate duration of key events is denominated by *t*. See the main text and Supplementary Information for the details.

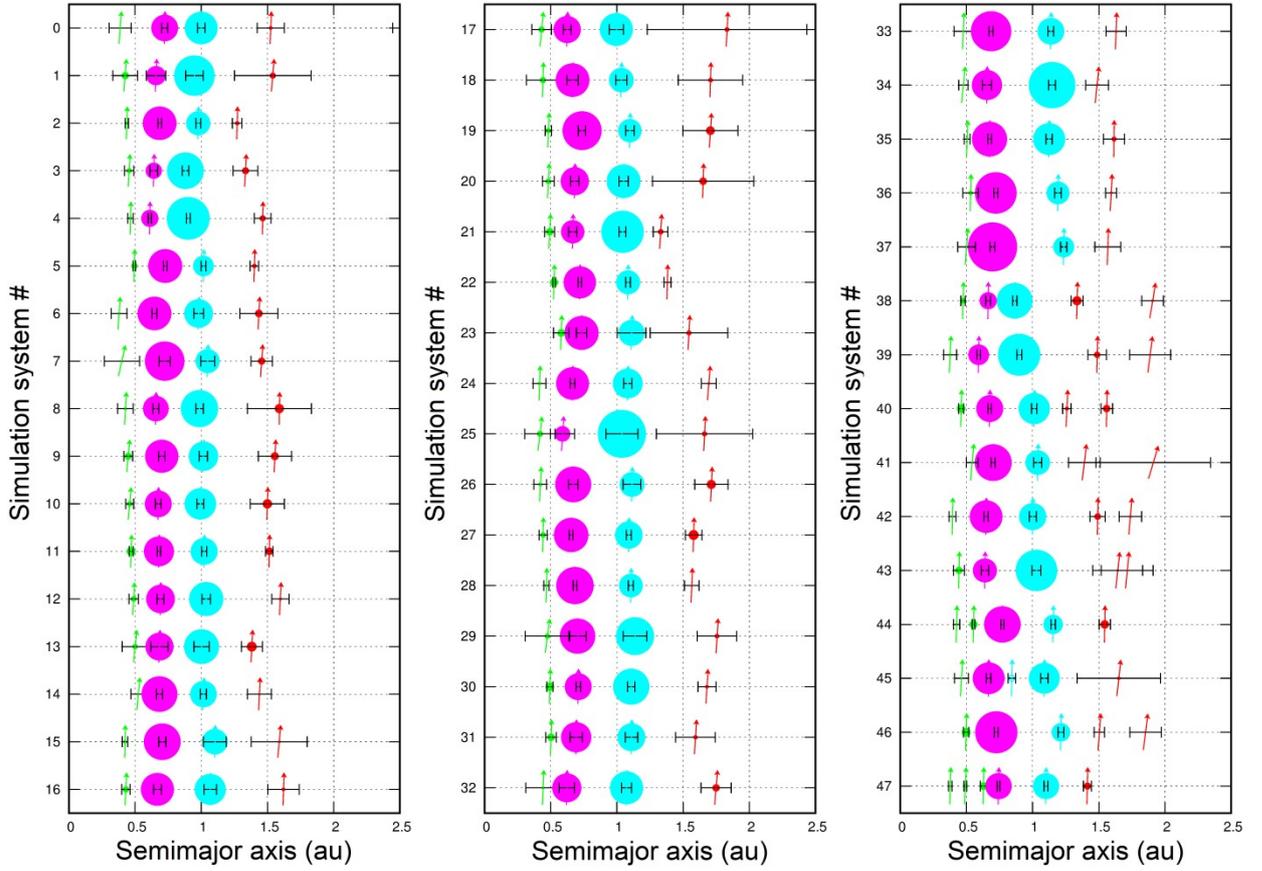

**Figure 2. Comparison of 47 individual analogue systems containing representative planet analogues of each terrestrial planet with the solar system planets (system #0).** Planet analogues of Mercury, Venus, Earth and Mars are indicated by green-, magenta-, cyan- and red-filled symbols, respectively. The inclination $i$ of the planets is represented by the angle between the vector and the perpendicular (e.g., the vector points to the top for $i = 0°$). The error bars indicate the variation in heliocentric distance based on the object's perihelion and aphelion. The radii of planet symbols scale in proportion to mass.

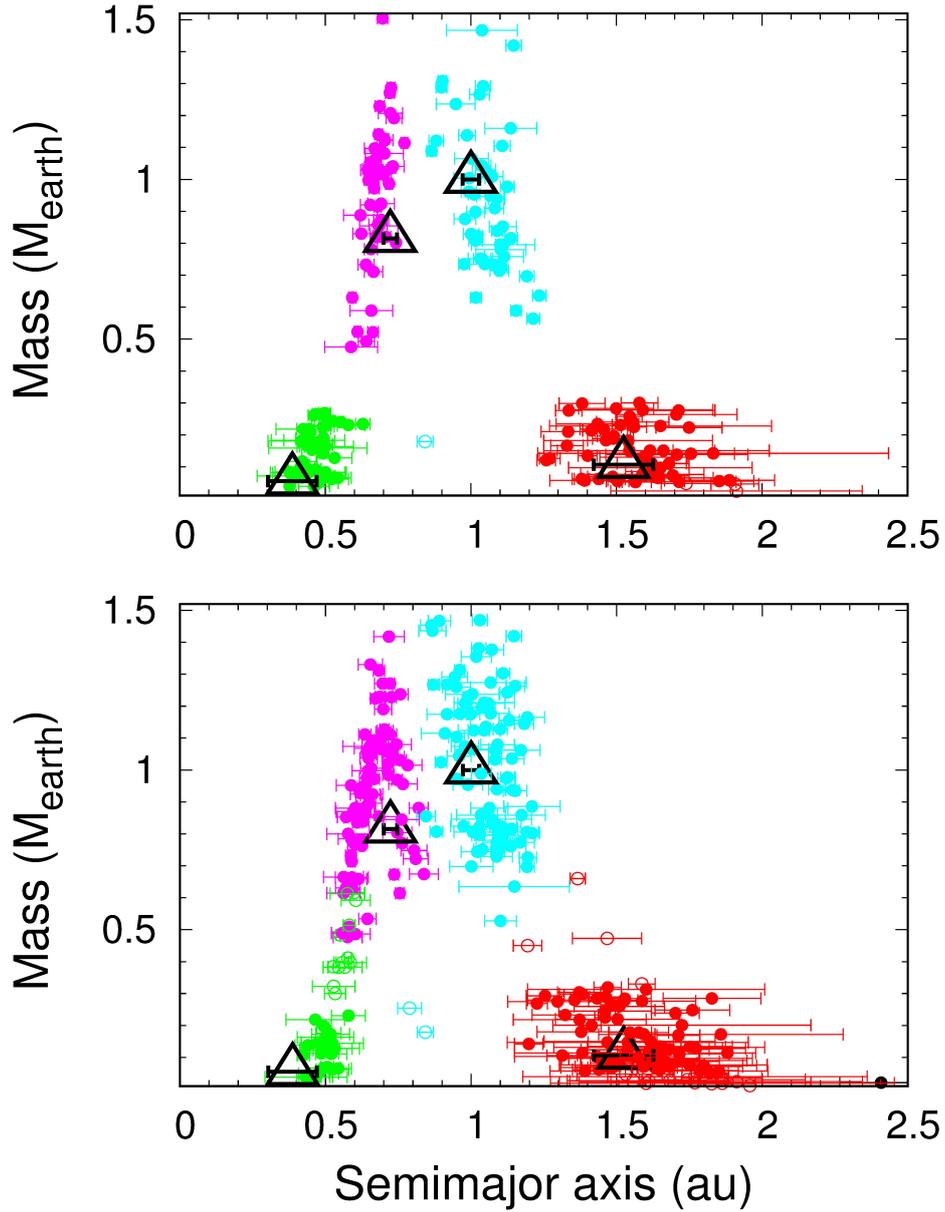

**Figure 3. Planets formed in 47 four-planet systems obtained from all simulations (top) and in analogue systems containing three or four representative terrestrial planets from simulations of the standard disk (bottom).** Planet analogues of Mercury, Venus, Earth and Mars are indicated by green-, magenta-, cyan- and red-filled symbols, respectively. The error bars indicate the variation in heliocentric distance based on the object's perihelion and aphelion. The large open triangles represent the terrestrial planets of the solar system.

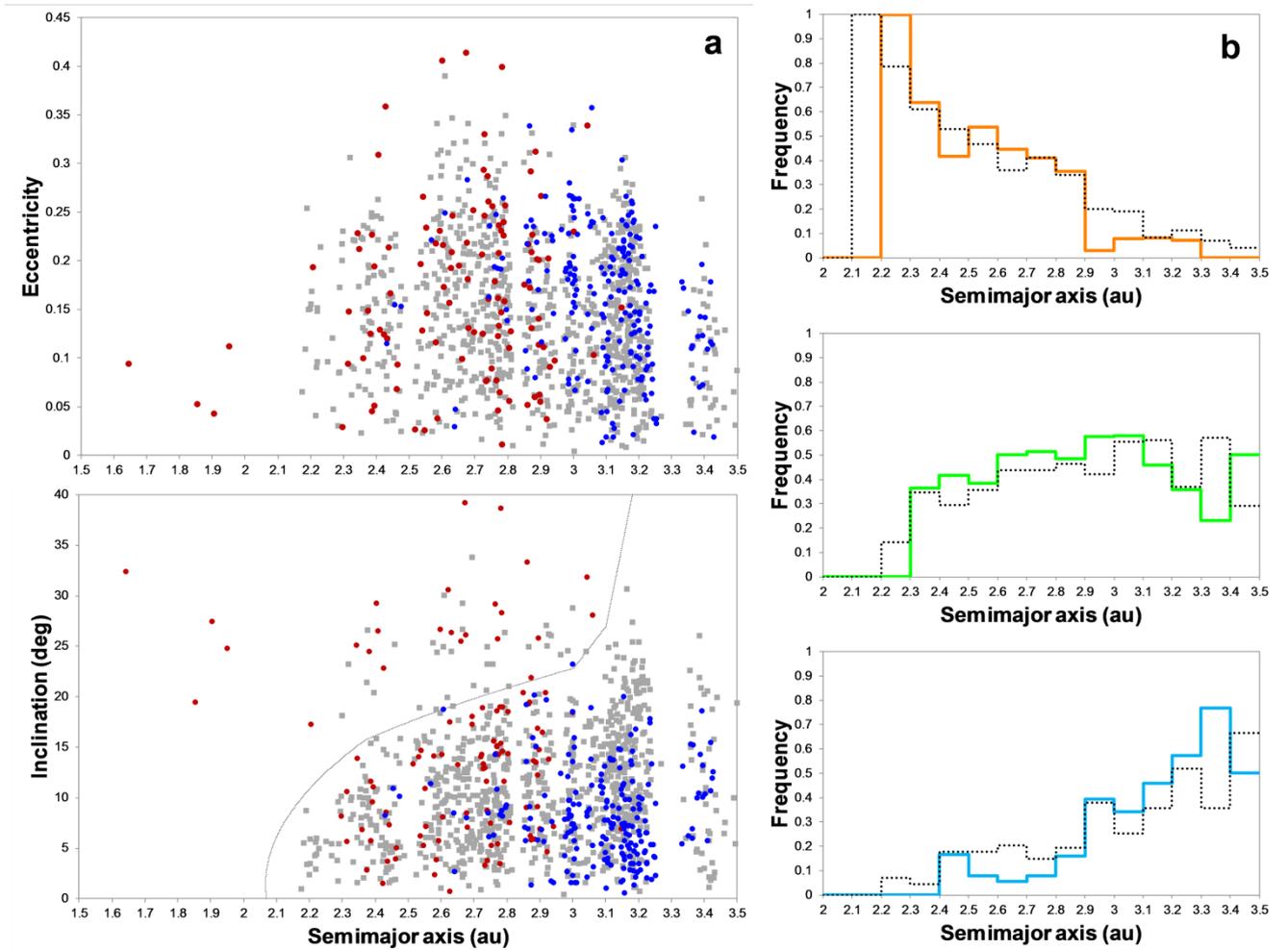

**Figure 4. Comparison of our representative asteroid belt constructed based on disk models that yielded terrestrial planet systems analogous to our own (coloured symbols/curves) with 895 large observed asteroids not belonging to asteroid families (greyscale symbols/curves).** Model asteroids represent the state after ~4.1 Gyr of dynamical evolution. Observed asteroids possess diameters D > 20–30 km (for albedos 0.1–0.2). **a)** The representative asteroid belt consists of local primordial asteroids in the protoplanetary disk at < 3.5 au (brown symbols) and asteroids captured from trans-Jovian orbits after giant-planet migration (blue symbols). The proportion of local:captured asteroids is 33%:67%. The curve indicates the location of the secular resonance nu6 (bottom). **b)** Both model and observed asteroids are classified as S- (top), C- (middle) and D/P-type asteroids (bottom). The histograms show the relative fractions of asteroid types per semimajor axis bin. The composition model considered the following proportions of S-, C-, D/P-asteroids: S80%-C20% (< 2 au) and S50%-C50% (> 2 au) for local and S5%-C47.5%-DP47.5% for captured asteroids. See Methods and Supplementary Information for details.

**Methods**

Basic assumptions

Our simulated systems started when the disk gas had already decayed, so its dynamical influence was negligible. The solar system nebular gas dissipated in ~4 Myr (ref.[49]), so for simplicity, we set 5 Myr as time zero in our scenario. At this time, the protoplanetary disk contained the newly formed giant planets, a small number of embryos, and many planetesimals[1,3,4,32,50]. Thus, following the methods used in several similar studies[1,3,4,5,6,9,19,21], we considered a disk containing embryos (lunar-Mars-mass objects), residual planetesimals and the Jupiter–Saturn pair with their current masses.

Jupiter–Saturn: pre-instability orbital properties

Hydrodynamic models of the early solar system when the disk gas was present found that Jupiter and Saturn can typically interact with their mutual 3:2 or 2:1 mean motion resonance (MMR)[51,52,53,54,55]. After the gas dispersal in the Jovian region, this orbital configuration was probably short-lived[23,56] and Jupiter, Saturn, and other giant planets in the outer solar system experienced a brief phase of dynamical instabilities[22,24,51] (henceforth 'instability'). Jupiter and Saturn then acquired their current orbits after a final phase of residual migration[24,51]. The post-instability evolutions of the Jupiter–Saturn 3:2 MMR orbital configuration have been explored in detail based on success metrics of fundamental aspects of the outer solar system[51]. However, recent work[57,58] showed that they might be better reproduced by a 2:1 MMR configuration with Jupiter and Saturn on moderately eccentric orbits (≤0.05). During the instability, close encounters with a Uranus- or Neptune-class icy giant may be needed to explain the current eccentricities of Jupiter and Saturn[51]. These results assume that Jupiter and Saturn had nearly circular orbits before the instability. However, at the end of gas dispersal, both planets may have inherited eccentric orbits[51,52,53,54,55,57,58], or acquired them by interactions with a self-gravitating disk[22] or massive primordial objects, as shown below.

Following this reasoning, we assumed that Jupiter and Saturn evolved to a 2:1 MMR acquiring non-zero eccentricities prior to the instability. The following evidence supports the hypothesis that both planets acquired a near-MMR configuration characterised by orbital period ratio PS/PJ ~ 2 and large libration amplitudes (~300-360°). First, a metastable near-MMR configuration is consistent with the well-supported idea that the instability occurred early in the solar system, as evinced by independent studies (see subsection below). Indeed, an early instability likely occurred if Jupiter, Saturn, and the other icy giants experienced such near-MMR states shortly after their formation. Second, the giant-planet formation was probably not a smooth process, and Mars-Earth-mass bodies or additional planets were present in the outer solar system. Collisions, gravitational encounters or secular effects involving such objects can turn a resonant configuration into a less stable near-MMR state[25,59,60]. Indeed, based on our database of giant planets locked in MMR chains, we confirmed such behaviour after performing auxiliary simulations including the presence of planetary bodies and the resonant Jupiter–Saturn pair, as similarly done in past representative work[25] (Supplementary

Information 3.5). These results agree with ref[25], which showed that this near-MMR state could arise after an initial 2:1 MMR Jupiter–Saturn pair suffers perturbations from such massive bodies. Furthermore, we found that the interactions of icy giants before the instability can lead to similar behaviour based on additional simulations considering Jupiter, Saturn, Uranus, Neptune and a fifth icy giant on mutually resonant orbits. Based on the auxiliary simulations, as illustrated by representative evolutions in Figures S5 and S6, our Jupiter–Saturn pairs exhibited moderately eccentric orbits, PS/PJ ~ 2 (with osculating variation) and chaotic orbital behaviour during our simulations' near-2:1 MMR phase, regardless of the initial instantaneous PS/PJ. Thus, any initial PS/PJ value close to 2 would work in our scenario. Overall, Jupiter and Saturn acquired averaged $e$ ~ 0.05-0.1 and evolved spontaneously in near-resonance with durations between a few to a few tens of Myr. Last, the tendency of exoplanets to exhibit orbital period ratios close to 3:2 and 2:1 MMRs suggests that near-MMR configurations are expected outcomes in planetary systems[61,62]. Therefore, we conclude that Jupiter and Saturn likely experienced near-resonant interactions in the 2:1 MMR on moderately eccentric orbits after gas dispersal (the starting time of our simulations).

Jupiter–Saturn: instability timing and post evolution

Evidence shows that the giant planets suffered dynamical instabilities less than 10–100 Myr after their formation[22,23,56,63,64,65]. However, as instability can excessively affect the terrestrial planets[66,67,68], it probably occurred in ~10 Myr or shorter timescales allowing Venus and Earth to acquire their current dynamically cold orbits via dynamical friction with remnant disk objects[69,70]. Here, we assumed that the Jupiter–Saturn pair interacted in the near 2:1 MMR configuration for a few Myr, after which the giant planets experienced instability. As the duration of instability is typically short[17,51,57,67] (< 100-a few hundred kyr timescales) and the post-instability residual planetary migration plays a minor role in sculpting the primordial asteroid belt beyond 2 au (ref.[17] and Supplementary Information 3.3 and 3.4), we placed Jupiter and Saturn on near-current orbits after the planets spent their initial stage of interactions in the near 2:1 MMR. This procedure has been used in studies of the long-term evolution of the asteroid belt[45,71]. It can avoid considering too many parameters related to the complex orbital evolution during the instability. In addition, from a terrestrial-planet formation standpoint, the details are probably not important when the disk is already mass depleted beyond ~1.5–2 au during the instability or the instability occurs early[6,9,33]. These conditions are consistent with our scenario. Nevertheless, the role of the instability (e.g., timing, weak vs strong, etc.) on the formation of the terrestrial planets is still poorly understood[31,72], so more detailed studies are warranted. Finally, the dynamical effects of the instability on the primordial asteroid belt are discussed in Supplementary Information 2.4 and 5.8.

Protoplanetary disks: fundamental properties

The terrestrial planets assembled by accretion of embryos and planetesimals within the protoplanetary disk[1,2,3,4,5,6,8,9,10,19,32,50,69,70]. In particular, the key factors determining the final terrestrial planets are the properties of the disk within 2 au, the fractions of disk mass represented by embryos or planetesimals and the orbits of the giant planets[3,6,19,20,50]. Here, we considered disks consisting of tens of embryos and several thousand planetesimals. Embryos concentrated from 0.35–0.5 to 1.15–1.5 au, while planetesimals were placed all the way to 3.5 au. At smaller distances, the disk typically contained more mass represented by embryos than by planetesimals. All disks started with a total mass equal to 1.9–2.1 ME within 1.2 au, usually needed to form Venus and Earth with their current masses. Planetesimals initially located at 2–3.5 au represented the primordial asteroid belt, the initial total mass of which varied between 0.5 ME and 2.2 ME to cover possible variations of a massive belt. There was no mass depletion initially in any region of the disk. Overall, it is worth noting that these initial conditions are consistent with the predictions of several models of embryo and planetesimal formation[26,32,34,50,70,73,74,75,76]. In this way, we kept the model simple by avoiding the problem of having to consider too many input parameters in embryo/planetesimal formation and gas dynamics (see also discussion in Supplementary Information 2.6 and 5.8a). A disk inner region probably played a fundamental role in forming Mercury[1,19,21,69], so we included this component in most of our disk models. We varied the disk inner edge (0.3, 0.4, or 0.5 au) and the initial total mass in the inner region to understand Mercury's formation better. Typically, the mass distribution in our disks resulted in a surface density that followed an increase from the disk inner edge until a distance threshold, then a decrease until the disk outer edge following a decay power law, as similarly modelled in the literature[1,4,6,9,35,77,78]. Our choice of a distance threshold at 0.8-0.9 au is supported by the findings that a narrow 0.7–1.0 au disk and disks with inner edges at 0.5 or 0.7 au generally produce Venus-like planets too close to the Sun[19,20]. We also tested mass concentrations within the disk core region, motivated by the need to explain the small Venus–Earth mutual distance. Such mass concentrations might result from planetesimal/embryo pile-up[11,12,13,14], so it is important to test their influence on terrestrial planet formation. Finally, our disk models' different surface density slopes/shapes reflect the need to satisfy the abovementioned constraints. All these variations in disk properties resulted in our distinct disk models (Figure S1 and Table S1). Embryos and planetesimals started initially on near circular and coplanar orbits ($e_0 < 0.01$ and $i_0 < 0.3°$). Thus, our primordial asteroids were dynamically cold at the start of the simulations.

Protoplanetary disks: object compositions

Inspired by asteroid taxonomy, we considered three major groups by composition in the disk: S, C and D/P. We assumed that our disk was composed of a mixture of local S-asteroids and C-asteroids. Local asteroids may have originated in situ or elsewhere (see below). In our nominal model, we assigned composition flags to these asteroids according to their initial location in the disk. This procedure resulted in proportions of C-asteroids of 10–20% and 50% within the < 2 au and > 2 au regions of our disk, respectively (conversely, 90–80% and 50% of S-asteroids within the referred regions). We chose a threshold of 2 au to define the terrestrial and asteroid belt regions. We also tested several taxonomic proportions but found that the nominal model yielded the best results (Supplementary Information 5.8). Some scenarios support this mixing of relatively "dry" volatile-poor S- and "wet" volatile-rich C-asteroids in the disk. For example, these primordial

C-asteroids could be objects implanted from the trans-Jovian region during giant-planet formation *before* instability[40,79]. This scenario predicts more significant contamination of implanted asteroids with increasing heliocentric distance beyond ~1 au. Another possibility is that the evolution of the water ice line at ~1-3 au during planetesimal formation contributed to the formation of C-asteroids beyond ~1-2 au (ref.[26,80,81,82]). Also, the high fraction of C types among asteroids at $a < 2.5$ au supports the idea that C-asteroids were present in the inner regions of the primordial asteroid belt[83]. Therefore, these pieces of evidence justify the above choices of an initial contribution of C-asteroids among the local population in the disk. The S-asteroids presumably formed in situ by the end of gas dispersal[16,32,40]. Primordial D/P-asteroids were likely captured from trans-Jovian reservoirs *during* the instability and migration of the giant planets[43]. Ref.[84] found that primordial Hilda asteroids and Jupiter Trojans were lost after the instability, so the currently known populations should consist of captured objects. As these populations consist of comparable C and D/P types, we considered that the captured asteroids were represented by 50% C- and 50% D/P-asteroids in our nominal model. In this way, the primordial asteroid belt was contaminated by both C- and D/P-asteroids by the time the giant planets acquired their current orbits. This contamination event is supported by dynamical modelling, spectral observations of asteroids and the presence of peculiar asteroids[43,85].

Concerning the chondritic compositions in the disk, we assumed that objects rich in enstatite and ordinary chondrites (EC and OC) were represented by S-asteroids and that the concentration of OCs increased for more distant asteroids. In addition, carbonaceous chondrite (CC)-rich objects were represented by C- and D/P-asteroids.

Protoplanetary disks: water mass fractions (WMFs)

We assigned distinct WMFs for the objects according to their initial locations in the disk to determine the amount of bulk water acquired by our terrestrial planets. Our goal was to constrain the water mass distribution in the disk by later identifying successful systems that satisfied the terrestrial planets' water constraint (Supplementary Information 1). In particular, we considered a wide range of WMFs for each investigated region at < 1.5 au, 1.5–2 au, 2–2.5 au and > 2.5 au: 0.001–0.01%, 0.001–0.5%, 0.1–10% and 5–40%, respectively. A total of 80 WMF models were investigated (Table S2). This exploration also allowed us to understand better whether these specific regions were water-poor or water-rich at the onset of terrestrial planet formation. Finally, these WMF distributions are consistent with models for the origin of water in the inner solar system[3,40,82,86] and the scenarios discussed in the previous subsection.

Main simulations

We performed 650 N-body simulations of terrestrial planet formation to investigate the disk models described above. We used an optimised version of the MERCURY integrator[87] to execute the simulations, including treatment of general relativity and calculation of the bulk density and size of the planets consistent with the terrestrial planets in the solar system[19]. The giant planets and embryos gravitationally perturbed one another. The planetesimals did not mutually perturb each other but gravitationally interacted with the planets and embryos. In these simulations, Jupiter and Saturn started with semimajor axes of $a = 5.5$ au and ~8.7 au and eccentricities of $e = 0.08$ and 0.1, respectively. An initial inclination of $i = 0.5°$ was assumed for both planets. For each disk model, we placed Jupiter–Saturn in 60 distinct configurations near their mutual 2:1 MMR according to the orbital period ratios PS/PJ = 1.97, 1.98 and 1.99 (20 configurations per PS/PJ). We uniformly varied Saturn's initial mean anomaly within 30–120°. All other angular elements were initially set to zero for Jupiter and Saturn. The initial resonant angles were 60–240°, but they quickly evolved to ~300–360° during the simulations. Here, our systems typically experienced the near 2:1 MMR Jupiter–Saturn for 5–10 Myr (20 Myr in a few cases), after which we took the orbital states of the embryos and planetesimals as the initial conditions for the next stage of the simulations with the giant planets placed on their near-current orbits. As expected after instability, we set the eccentricities slightly above the current values for Jupiter and Saturn. This procedure also allowed the eccentricities to damp slightly to current levels via dynamical friction with remaining embryos/planetesimals. Then, we evolved 50–100 systems per disk model until the total time reached 400 Myr, representing the system's post-instability evolution. Overall, the PS/PJ ratio remained slightly below the observed 2.49, so the influence of the near 5:2 MMR was negligible. Additional auxiliary simulations are described in Supplementary Information 3.

To investigate water delivery to the terrestrial planets, we considered 87 systems that contained planets with optimal global properties (i.e., the median orbits and masses over all analogues of each planet were within 10% and 25% of the respective actual values) at the end of 400 Myr based on disk models A, B and C, which together represented the 'standard disk'. As the main results were similar among all disk models tested in this study, the selected systems should be good representatives of the inner solar system (Figure 3).

We did not include fragmentation in our simulations, but this was an acceptable assumption. First, simulations of the terrestrial-planet formation including fragmentation yielded system outcomes similar to those that did not[10,33,50,88,89,90] because the generated fragments can be re-accreted by the forming planets. Also, it is unclear if fragmentation plays an important role or not in explaining the current orbit and mass of Mercury[91]. Furthermore, adding more confusion to the picture, the simulations that implemented fragmentation yielded more massive Mars analogues, thus decreasing the success of Mars formation[33]. The stochastic nature of planet formation, the different techniques/codes used to implement fragmentation in simulations, and the uncertainties regarding the fragmentation of primordial solar system objects (e.g., critical impact energy per unit mass $Q_*$) make it difficult to judge and compare the results of past studies regarding fragmentation. Briefly, the role of fragmentation in terrestrial-planet formation remains a matter of debate, and further detailed studies are warranted. Nevertheless, a common observation in studies that included fragmentation is that dynamical friction was enhanced, which could lead to dynamically colder final planets[33,89]. Nevertheless, dynamical friction operated in our simulations because we used a large number of planetesimals that decayed over tens of Myr.

The simulation time step was 4.565625 days for disks A, B, C, D, E and 3.6525 days for disks

Ia–e to ensure reliable calculations for planets with orbits similar to Mercury. Objects that evolved to heliocentric distances < 0.15 au or > 20 au were discarded from the simulations. This procedure did not influence our main results because these extreme objects can collide with the Sun or be gravitationally ejected by Jupiter on very short timescales.

Asteroid belt formation: model, simulations and observations

We investigated the origin and evolution of the asteroid belt by considering a representative asteroid belt obtained in additional long-term simulations. Residual migration was not considered in this investigation, as justified in subsection 'Jupiter–Saturn: instability timing and post evolution' above and in Supplementary Information 3.3. Due to computational constraints, we limited this investigation to remnant asteroid belts obtained in the systems of the standard disk, which initially contained a primordial asteroid belt, as described above. First, we selected 202 systems containing three or four terrestrial planet analogues after 100 Myr of post-instability evolution from the main simulations described previously (100 Myr was the closure time of planet formation in this investigation; henceforth 't100'). Thus, only systems that formed good representatives of the inner solar system were considered in this investigation. We further selected 46 systems with asteroid belts depleted at levels < 5% in which > 50% of the local asteroids acquired $i > 10°$. This procedure identified asteroid belts that experienced substantial depletion and $e$-$i$ excitation simultaneously. Then, we obtained the orbital states of local asteroids remaining in those systems at t100. Next, we considered asteroids captured in the asteroid belt from trans-Jovian orbits based on data obtained in giant-planet migration simulations[44]. These captured asteroids acquired their final orbits within ~4 au after interacting with Jovian mean-motion and secular resonances[43,44,172]. Notably, the orbital structure of the captured asteroids and their capture efficiency in distinct regions of the asteroid belt were quite similar to those found by a distinct model that included the instability[43], so the results of that study[44] provide an acceptable representation of such a captured population at the end of giant planet instability/migration. We then obtained the orbital states of our captured asteroids after evolving them to t100. In the last stage, we combined the local and captured asteroids obtained at t100 and evolved them for a further 4 Gyr under the gravitational influence of the eight solar-system planets and the four most massive asteroids (Ceres, Vesta, Pallas and Hygiea). In this investigation, assuming that the planets were fully formed and placed on their current orbits allowed us to investigate the long-term evolution of the asteroid belt accurately because the associated mean motion and secular resonances in the inner solar system were correct. We found that 106 local and 658 captured asteroids survived within 2–3.5 au at the end of this stage (out of ~78k local and 2118 captured asteroids, respectively, where the latter objects were captured out of 3 million trans-Jovian objects[44]). Finally, by random sampling of the captured population, we built representative asteroid belts based on relative contributions of 33%:67%, 50%:50%, and 67%:33% for local and captured asteroids, respectively (Figures 4, S7 and S8). We also tested 80%:20% and 20%:80% proportions and found that in both cases, the resulting representative asteroid belts showed strong peaks in the semimajor axis that were inconsistent with actual observations. Therefore, the contribution of local and captured asteroids in our adopted representative asteroid was limited to a factor of 2 (i.e., comparable populations).

To compare our asteroids with observations, we selected 895 large asteroids with absolute magnitudes of H < 11 (> 20–30 km for assumed albedos 0.1–0.2) and not belonging to any asteroid family on 7 April 2021 from the AstDyS database. These asteroids are large enough that long-term collisional and non-gravitational effects would not affect their orbits[6]. In addition, the discovery of asteroids with H < 11 is essentially complete[92]. The compositional taxonomy of S-, C-, and D/P-type asteroids is based on the Bus-DeMeo classification[15,93].

Terrestrial planet system classification

It is crucial to identify the analogues of Mercury, Venus, Earth, and Mars in systems obtained in terrestrial-planet-formation studies[19,20]. This procedure also allows us to properly identify analogue systems that contain three or more planet analogues. As discussed in detail in ref.[19], using a proper classification scheme can also mitigate the problem of incomplete and misclassification, thus reducing the chances of reaching misleading conclusions. This study used ref.[19]'s rigorous classification algorithm to identify all planet analogues formed in a given system. The algorithm obeys the following main steps: typically, the two most massive planets are identified as the Venus and Earth analogues of the system; next, the Mercury and Mars analogues are identified as the planets formed adjacent interior and exterior to the Venus and Earth analogues, respectively; finally, systems not analogous to the inner solar system are discarded (e.g., a system with a massive planet [> 0.32 ME] exterior to the Mars analogue). The planetary mass ranges considered for analogue candidates were 0.025–0.27 (Mercury), 0.4–1.5 (Venus), 0.5–1.5 (Earth) and 0.05–0.32 ME (Mars). Several past studies used similar mass ranges[9,13,21,94]. The upper limit for Mercury is slightly larger than usual to allow the possibility of mass-loss via erosive cratering[95] or hit-and-run collisions[91], which are not modelled here (see also Supplementary Information 5.2). Nevertheless, the influence of minor changes in mass ranges is unimportant compared to other model parameters, such as disk properties and the giant planet orbits[20]. A planet was defined as any object with a mass m ≥ 0.025 ME (a minimum of 50% of the mass of Mercury). Our classification algorithm required our terrestrial planet systems to contain planets analogous to the Venus–Earth pair and Mercury or/and Mars in the same system. In total, we identified 221 terrestrial systems that satisfied the above conditions. Then, we examined the properties of these analogue systems and the planets formed therein against fundamental constraints in the inner solar system (Supplementary Information 1). The numerical results discussed in this article are medians or ranges of medians obtained from our combined disk models (unless expressly stated otherwise): standard disk (ABC), representative disk with small inner regions (A–E) and representative disk with extended inner regions (Ix) (Tables S1, S4 and S5). Finally, we note that the influence of our system/planet classification details on our results was unimportant. In general, the choice of Earth analogue mass classification (e.g., using a stricter 0.8–1.2 ME), system type (3-P or 4-P) and disk sample (individual or combined group) had little influence on the main properties of our terrestrial systems and their planets. Furthermore, the main results related to our planet were also insensitive to the classification details of our Earth analogues.

**Data availability**

The main findings of this study are supported by the data presented and the Supplementary Information. Additional data can be obtained from the corresponding author on reasonable request. The asteroid orbital data are available from the AstDyS database at https://newton.spacedys.com/astdys/. The asteroid taxonomy data based on the Bus-DeMeo classification system are available at http://www.mit.edu/~fdemeo/publications/alluniq_adr4.dat, https://sbn.psi.edu/pds/resource/taxonomy.html and https://sbn.psi.edu/pds/resource/busdemeotax.html.

**Code availability**

MERCURY is publicly available at http://ascl.net/1201.008. The optimised MERCURY used in this study is based on the tweaked version of the code available at https://gemelli.spacescience.org/~hahnjm/software.html.

**Acknowledgements**

We are grateful to all five referees (Matthew Clement and four anonymous) for several helpful comments, which allowed us to improve the overall presentation of this work significantly. We also thank F. E. DeMeo for clarifications about the asteroid taxonomy data and J. P. Greenwood for commenting on the terrestrial planets' water contents. The simulations presented here were mainly performed using the general-purpose PC cluster at the Center for Computational Astrophysics (CfCA) in the National Astronomical Observatory of Japan (NAOJ). We are grateful for the generous time allocated to run the simulations.

**Author contributions**

P.S.L. designed the simulations and wrote the paper. All authors performed the simulations, analysed the data, and discussed the results thoroughly.

**Competing interests**

The authors declare no competing financial interests.

# Supplementary Information

## 1. INNER SOLAR SYSTEM CONSTRAINTS AND SUCCESS CRITERIA

A successful model for the inner solar system should satisfy all the constraints described below.

*Formation of Mercury, Venus, Earth and Mars analogues.* Reproducing the orbits and masses of the four terrestrial planets simultaneously remains elusive[1,2,4,5,6,13,19,20,21,32,33,69,70,94]. Given this difficulty, an analogue system has to form a minimum of three planet analogues, including the Venus–Earth pair. In addition, while the orbits of Venus and Earth are dynamically cold ($e \sim 0.03$ and $i \sim 2°$), Mars has a mild orbit ($e \sim 0.07$; $i \sim 4°$). Finally, Mercury's excited orbit at $a \sim 0.4$ au ($e \sim 0.2$; $i \sim 7°$) and its low mass are difficult constraints to satisfy among the terrestrial planets[19,21,96].

*Formation timescales of the terrestrial planets.* Earth's formation timescale is often assumed to be that of the formation of the Moon, but the latter is not well constrained (see below). Isotopic constraints indicate that 80% of Earth's mass most likely accreted within a timescale of 10–35 Myr, while the remaining mass accreted over the next tens of Myr[42]. Venus probably experienced a similar accretion history. The bulk of Mars' mass (>80–90% of the final mass) could have accreted in less than 10 Myr according to Hf-W-Th chronometry[97]. However, the same isotopic data is also consistent with longer accretion timescales if Mars formed protractedly involving giant impacts[98]. Indeed, some studies supporting the latter idea found Mars' formation timescales of ~15–20 Myr (ref.[99,100,101]) or even ~60 Myr (ref.[102]). Here, consistent with these estimates, a timescale of ≤ 15 Myr was assumed for Mars formation. Timescales up to 50% longer (~23 Myr) were acceptable, provided that ≥ 80% of the mass of Mars was accreted within 5 Myr. Lastly, to our knowledge, Mercury's formation timescale is unconstrained.

*Formation of the Moon.* Estimates of when the Earth experienced the giant impact that formed the Moon varied within $t = 25$–245 Myr in the time framework of our model[36,37,103,104,105,106,107]. This wide range of timescales highlights that Moon formation is still a matter of debate, which includes a variety of plausible impact geometries, thermal states, compositions, and masses of the impactor and the target[106,107,108,109,110,111,112,113,114,115,116]. While it is typically assumed that a ~Mars-mass object caused this impact[108], qualified Moon-to-Mars mass impactors are as capable of forming the Moon[117], which is supported by Moon-formation models[111,112]. Geochemical constraints limit the Moon-forming impactor to at most 0.1–0.15 ME and favour less massive impactors[118]. An independent analysis concluded that one or a few impactors with masses within 0.01–0.1 ME formed the Moon[119]. Here, we considered these uncertainties and assumed that impactors with > 0.02, > 0.03, > 0.05 or > 0.10 fractions of the target's mass were representative of Moon-forming giant impacts.

*Late veneer mass delivered to Earth.* The fraction of mass delivered to the Earth's mantle via impacts of remnant objects during late terrestrial-planet formation was within 1% of Earth's mass[47,48,117]. In addition, most of the objects that contributed to the late veneer were probably rich in dry enstatite chondrites[41,120,121]. Finally, more massive impactors (≥ 0.001 ME) likely deposited only a fraction of their masses into Earth's mantle. In particular, for 0.01- and 0.03-ME impactors, the contributions were only 10–30% and 10–20%, respectively[122]. This effect was considered in our simulations by considering the contribution upper limits conservatively.

*Origin of water on Earth and the other terrestrial planets.* Earth's bulk water was likely delivered to our planet by impacts of various water-bearing objects present in the protoplanetary disk during its formation[3,4,19,38,39,40,41,123]. Estimates of the mass of water on Earth vary significantly[41,103,124,125]. Here, Earth analogues were required to acquire water mass fractions (WMFs) in the range $5–25 \times 10^{-4}$, or 2–10 times Earth's current water in the oceans. The WMFs of Venus and Mars are much less constrained. Here, we assumed a minimum WMF of $1 \times 10^{-5}$ and a maximum WMF of $5 \times 10^{-4}$ ('dry Venus') or $5 \times 10^{-3}$ ('wet Venus') for Venus. For Mars, we assumed that a range of WMFs within $0.5–20 \times 10^{-4}$ was required[41,126,127,128,129]. We considered a WMF model successful if the three median WMFs obtained for the analogues of Venus, Earth, and Mars simultaneously satisfied the estimated WMF intervals for each of the planets (Table S2). Finally, 5–30% of Earth's water was likely delivered after the Moon-forming giant impact, so the bulk of the Earth's water was accreted before this event[41].

*Terrestrial planet system properties.* The angular momentum deficit (AMD) measures the dynamical excitation of a system via the eccentricities and inclinations of planets formed therein. At the same time, the radial mass concentration (RMC) evaluates the mass distribution of the planets in terms of their orbits (see ref.[1] for more details). Here, consistent with representative studies[4,6,9], the final systems should possess a maximum of 200% of the AMD and 50–200% of the RMC of the terrestrial planets.

*Absence of planets in the asteroid belt.* Whether planets can form locally or move in from other regions, they must not acquire stable orbits in the asteroid belt after terrestrial-planet formation; otherwise, peculiar orbital features that do not match observations arise in the belt[130].

*Orbital architecture, compositional taxonomy and the low mass of the asteroid belt.* A successful asteroid belt model should simultaneously explain the orbital/compositional structure and the low mass of the asteroid belt at 2–3.5 au (Figure 4). Our sample of observed asteroids is a good representative of the asteroid belt (Methods) and exhibit *a-e-i* distributions consistent with independent studies of the belt[131,132]. In particular, an essential constraint regarding inclinations is the ratio of asteroids with inclinations above the location of the nu6 secular resonance in *a-i* space, which is currently $f6 = 0.07$ at $a < 2.82$ au. About taxonomy, asteroids are often classified as S- and C-type asteroids, which are assumed to be poor and rich in water (or volatiles), respectively[123]. However, studies of surface/geochemical processes and spectral taxonomy indicate that while C- and D/P-type asteroids would be water-rich, they likely belong to distinct groups[15,93,133]. Finally, while S- and D/P-type asteroids are concentrated in the inner-middle and outer regions of the belt, respectively, C-asteroids seem to be broadly present throughout the entire belt. Another fundamental constraint is the total mass of the asteroid belt, estimated at $5 \times 10^{-4}$ ME (ref.[15]).

# 2. CHALLENGES IN MODELS OF TERRESTRIAL-PLANET FORMATION: A CRITICAL OVERVIEW

*2.1 Classification of terrestrial planets and planetary systems.* The lack of proper identification of planet/system analogues and oversimplistic classification schemes are overlooked in terrestrial-planet-formation studies[19]. For instance, many representative models discussed below show the planets mixed without discriminating planet analogues or individual planets without clarifying if they belong to analogue systems. We believe that analysing individual planets (e.g., constraints related to Earth) and terrestrial planet systems should be based on adequately classified systems containing three or four analogues of the terrestrial planets. However, only a few recent models have considered the importance of classifying analogue systems in their analysis[6,9].

*2.2 Is the narrow disk 0.7–1.0 au an adequate baseline?* The outcomes of narrow protoplanetary disks with mass concentrated at 0.7–1.0 au (ref.[7,10]) (henceforth 'canonical annulus') have served as the basis for several Mars formation models: Grand Tack[8,117,124,134], early giant-planet instability[6,9], empty asteroid belt[18], pebble accretion[32,135] and sweeping secular resonance[136]. In particular, the Grand Tack and early instability models have explored the Mars formation problem in detail. Although narrow disks obtained by these models appear to solve the Mars small-mass problem, there has been insufficient discussion about other outstanding problems. To cite a few, simultaneously forming Mercury (in addition to Venus/Earth or Mars in a given system), explaining the cold and close-in orbits of the Venus–Earth pair, reproducing the moderately excited orbit of Mars and its mutual distance with Earth, and confronting other fundamental constraints (Section 1) in systems that contain three or more planet analogues with the observations in those models. More importantly, detailed investigations of three- or four-planet analogue systems properly identified in such disks indicate that they have difficulties reproducing the terrestrial planets and satisfying other constraints in the inner solar system[19,20].

First, after applying the same classification system used here to the best representative canonical annulus model (50 systems of *D1* disk modelled in ref.[20]), we found that the formation of Mercury analogues and systems simultaneously containing analogues of the four terrestrial planets (henceforth '4-P systems') were inefficient at 6% and 4%, respectively. In contrast, here (disk Ix), these efficiencies reached 34% and 13%, respectively. The problem of forming Mercury in tandem with the other terrestrial planets is a well-known issue (Section 2.3). Furthermore, for several types of similar narrow disks (e.g., akin to Grand Tack) tested in our previous work[19,20] (e.g., *T7–10* and *T7–12* narrow disks with core mass at 0.7–1.0(1.2) au + low-mass components beyond 1.0(1.2) au and *D1* narrow disks) the analogues of Mercury formed too far from the Sun typically at semimajor axis $a > 0.52$ au and were located too close to the Venus analogues. Indeed, using the same success criterion $0.67–1.33 \times$ dMV (Mercury–Venus mutual distance) as used here, we found 0% success for *D1* (ref.[20]), *T7–10* and *T7–12* disks[19]. In contrast, we obtained 30% in our scenario (Section 5.2). Similarly, the analogues of Mars formed too close to the Earth analogues in *T7–10* and *T7–12* disks (0.32 and 0.39 au, respectively). This problem was not observed in our results. Furthermore, analysis of planet analogues in orbital-mass space revealed the failure of the canonical annulus model to reproduce simultaneously the orbits and masses of the Venus–Earth pair[20] (that work's figure 2 supports this conclusion). Another issue is that concentrating disk mass within a 0.7–1.0 au annulus yielded a low probability of forming Venus and Earth within the same boundaries. Because the disk expands due to self-stirring[6], Venus and Earth tend to form too far from each other. Despite differences in model details or classification schemes, representative studies[6,9,31,33,134] of the above scenarios reveal similar trends regarding Venus/Earth. In addition, the Earth analogues generally acquired 90% of their final masses too fast: ~7-16 Myr (medians), consistent with the Grand Tack's results. Consequently, these planet analogues experienced Moon-forming giant impacts too early, with success fractions of 0%, 13% and 33% for *T7–10*, *T7–12* and *D1* disks, respectively (For an ITR > 0.05), and accreted too much late veneer mass after these impacts in agreement with the trends found in ref.[117,124,134,137].

Finally, the summaries of formed planets as illustrated in plots of orbit vs mass reveal potential systematic issues in the studies mentioned above: 1. Overly noisy or peculiar distributions of planetary orbits/masses (e.g., Venus-like planets concentrated at ~0.5 au); 2. Recurrent presence of undermassed Venus-/Earth-like planets or overmassed Mercury-/Mars-like planets; 3. Lack of identification of individual planet analogues. These points contrast with our improved results regarding orbit-mass distributions of the planet analogues (Figure 3).

*2.3 Forming Mercury and 4-P systems.* The formation of Mercury is an outstanding problem that has been largely neglected in terrestrial-planet-formation studies[2,3,4,5,8,10,12,13,27,32,33,35,50,69,70,77,78,88,90,94,136,138,139,140,141,142] (see also discussion in ref.[19,20]). Other studies have not obtained satisfactory results[1,6,7,9,11,18,31,72,89,134,137,143]. Although recent dedicated models have explored different scenarios for Mercury's formation in terms of its orbit, mass and internal structure[21,91,96,143,144], there is still no consensus on the best model regarding the reproduction of Mercury's main properties. More detailed studies are warranted to investigate whether any of these scenarios could form 4-P systems and satisfy other constraints in the inner solar system.

Because distinct classification schemes were employed and other details were lacking in past work, a direct comparison with our results is problematic. Nevertheless, we briefly discuss a few results below for completeness. We limit the discussion to the terrestrial planet formation models that considered Mercury formation in their terrestrial systems. First, Mercury-like planets were obtained in ≤1% of the systems in the Grand Tack[134,137] and the early instability models[9,33]. More recent investigations of the early instability model yielded ~4-6% and <1% probability of producing Mercury-like planets in Mercury–Venus systems[31,91] and 4-P systems[91], respectively. Despite the above caveats, these values are lower than those obtained in our best disk models: 34% and 13% for Mercury analogues (in Mercury–Venus–Earth and 4-P analogue systems combined) and 4-P systems, respectively.

*2.4 The influence of Jupiter–Saturn evolution during the instability.* The early instability model considered instabilities valid for orbital period ratio PS/PJ as large as 2.8 in their first series of studies[9,16,33]. However, because secular resonances associated with Jupiter and Saturn can strongly perturb the asteroid belt, it is unclear if instability evolutions characterized by PS/PJ ~ 2.5 (interacting with the 5:2 MMR) or crossing of the 5:2 MMR (for PS/PJ = 2.5-2.8) perturbed the belt unrealistically. Ref.[6] considered three individual instability evolutions characterized by PS/PJ ~ 2.3, ~2.45, and ~2.45-2.55. The latter case was also used as the basis for a representative asteroid belt formation model[17]. However, this particular instability evolution raises concerns about the influence of the 5:2 MMR. Our results suggest the intriguing possibility that a similar Jupiter–Saturn chaotic excitation (JSCE) mechanism based on the near-5:2 MMR played a role in perturbing the asteroid belt in that study. On the other hand, even if this were the case, one would need to explain how Jupiter and Saturn experienced PS/PJ ~ 2.5 and then later acquired their current smaller PS/PJ (~2.49) that is not experiencing JSCE. Furthermore, the terrestrial systems obtained under this specific instability evolution are also problematic: the planets seem dynamically too excited, and Mars-like planets are scarce[6]. Our scenario does not suffer from the shortcomings mentioned above. We also confirmed that the influence of the instability was modest in our investigations, so our main results stand (Methods and Section 3).

*2.5 Explaining the orbital distributions and compositional taxonomy of asteroids.* Asteroids obtained in Grand Tack[8,134], early instability[6,9,16,33] and asteroid belt formation models[17,45,145] are typically overexcited in inclinations compared to observations. This issue is closely related to the longstanding problem of reproducing f6 (Section 1), which remains unexplained by these models. In fact, models that consider the Grand Tack's results at gas dispersal as their initial conditions recognise the overexcitation problem, so they assume instead that the asteroids concentrated at $i < 20°$ before the instability[6,45,145]. Although the final phase of Jupiter–Saturn residual migration could reduce f6 by 60% on average, the resulting f6 values were eight times the observed one in the early instability scenario[146]. Other issues reported in some of those studies are the deficit of low-$e$/low-$i$ asteroids[17,45] and overexcitation of eccentricities[84], which are also seen in asteroid belts[6] obtained for the instability evolution used in ref.[17]. The most promising asteroid belts obtained in ref.[6] are based on the instability evolution characterised by PS/PJ ~2.45. However, in this particular case, the terrestrial systems appear to possess planets too dynamically excited, overly massive Mars-like planets, and Venus-like planets concentrated at ~0.5-0.6 au. In contrast, in the framework of the JSCE scenario, our asteroid belt acquired a less excited orbital structure more compatible with observations. Another representative scenario is the empty primordial asteroid belt model[18]. However, the lack of details in that study precludes a more in-depth comparison of model results with the various properties of the asteroid belt (Section 1) or discussion about the influence of the giant planets' orbital evolution. Contrary to the long-term investigations in our study, the obtained asteroid belts in the early instability and empty asteroid belt were not evolved to the age of the solar system (e.g., ~4 Gyr), making a comparison with observations less straightforward. Finally, the compositional structure of the asteroids is seldom addressed with detailed statistics. The Grand Tack model considered S- and "C-type" populations (this population mixed C- and D/P-type objects), but their distributions differ from the current observed one (Figure 4b). Similarly, the early instability model found their S- and C-type populations concentrated in the inner and outer regions of the asteroid belt. However, these results were contaminated by systems that experienced too violent giant planet instabilities. Here, we argue that analysing three main populations (S-, C-, and D/P-types) is more comprehensive and can lead to a better understanding of asteroid belt origin and evolution (Methods and Section 5.8).

*2.6 Early formation of terrestrial planet systems.* Recent planetesimal/embryo-formation models have reported forming terrestrial planets[11,13,27]. Although these models have yet to demonstrate that their results can satisfy the various constraints of the inner solar system, we discuss them here for completeness. First, these models deal with a large number of parameters related to gas dynamics, pebble accretion or planetesimal/embryo formation. For this reason, the parameters are explored preferentially to produce disk mass concentration near Venus–Earth's current orbits or "empty" asteroid belts to prevent overmassed Mars-like planets from forming. Second, the formation and dynamical evolution of Jupiter–Saturn played an essential role during the disk gas phase, but the giant planets were not included in representative models (1) and (2) discussed below. We now discuss three representative models.

(1) The strength of the embryo-convergence model[11] is the tendency of Venus- and Earth-like planets to form mutually close within the convergence region. However, this orbital convergence is also a problem because Mercury- and Mars-like planets tend to form too close to their Venus and Earth counterparts. Regarding Moon formation, most of the obtained Moon-forming giant impacts occurred too early, with a median probably falling within 10-20 Myr after gas dispersal. Despite the low likelihood of late impacts, it is assumed that a 5-planet system experienced one to match the Moon-formation constraint, thus resulting in a 4-planet system. This hypothesis requires that the 5-planet system was dynamically unstable at some point after its formation. However, the orbital excitation experienced by the planets during this late terrestrial planet instability was not discussed. The eccentricities of the Venus/Earth-like planets were generally ~0.015-0.02 before this event, so both the terrestrial and the giant planet instabilities are likely to overexcite the orbits of Venus and Earth[66,67,147,148]. It is also difficult to evaluate the level of success of the individual planets or systems formed because the planets are shown mixed, and the classification of analogue systems is unclear. Finally, the information about the asteroid belt is insufficient to allow an adequate comparison with observations.

(2) The pebble accretion model[27] relies on two key assumptions: i) formation of the building blocks (planetesimals) of Venus, a 0.6 ME-Earth, Theia (the Moon-forming impactor), and Mars at the ice line location fixed at 1.6 au; ii) a 0.4 ME-Theia located at ~1.2-1.5 au after gas dispersal. However, the ice line likely wandered around 1-3 au during planetesimal formation[26,80,81,149]. Thus, this would lead to various possible growth tracks decreasing the likelihood of the conditions described in i-ii) above. Also, the likelihood of Theia colliding with Earth late enough and the level of orbital excitation acquired by Venus, Earth, and Mars after terrestrial *and* giant planet instabilities are unclear. Furthermore, Mercury's formation is difficult to evaluate, as the planet was not explicitly modelled in this scenario. Finally, a model of asteroid belt formation was not presented.

(3) The planetesimal-ring model[13] claims that the terrestrial planets formed from a disk containing

~2.5 ME of mass concentrated at ~0.7-1.5 au. Such disks are wider than the 0.7-1.0 au narrow disks discussed above, resulting in worse final terrestrial-planet systems in orbit-mass space. In particular, this is evinced by the formation of Venus-/Earth-like planets covering a similarly wide range at $a \sim$ 0.5-1.5 au and the existence of individual systems containing three or more such massive planets. Another piece of evidence is the presence of overmassed Mars-like planets. Thus, the chances of obtaining low-mass Mars-like planets are more diminutive in spatially wider disks. Mercury formation was also neglected. Furthermore, there are some unclear points about the classification scheme. Which planets were explicitly classified as the individual Venus, Earth and Mars analogues? How additional Venus-/Earth- or Mars-like planets were classified in specific systems? Can a system with three or more Venus-/Earth-like planets be considered a good representative of the real one? Finally, there is an insufficient discussion about the formation of the asteroid belt, so it remains to be shown if the belt's orbital structure, compositional distributions, and low mass can be explained consistently in that scenario.

Concluding, any successful model for the inner solar system should reproduce the four terrestrial planets and the main properties of the asteroid belt in a single scenario. It is also clear that the initial conditions of the disk from which the inner solar system originated must be better constrained. Despite the limitations of our model, we believe that it promotes advancement in our understanding of terrestrial and asteroid belt formation.

3. AUXILIARY SIMULATIONS AND RESULTS

*3.1 Control systems.* We simulated 70 control systems that considered Jupiter and Saturn on their current orbits from the start using disks A, B, C, D and E (Table S1). A median of 35% of asteroids survived in these systems after 400 Myr of evolution. Furthermore, as there was no effective mechanism to excite the inclinations, all the asteroids still had low inclinations ($< 5°$) at the end of the simulations. In addition, even the best control systems formed Mars analogues systematically more massive than Mars, with orbits too close to Earth analogues and on timescales longer than 25 Myr. Therefore, a system starting with the current giant planets cannot explain the formation of Mars and the asteroid belt for a massive disk extending beyond 2 au. This result strengthens the case that either the disk was dynamically truncated early or the primordial region beyond ~1 au was never high-mass.

*3.2 Jupiter–Saturn 2:1 MMR systems.* We also simulated 120 systems containing Jupiter and Saturn deeply locked in a 2:1 MMR (i.e., with small resonant angle amplitudes) using disks A, B, D and E. We found no dynamical excitation beyond ~1–1.5 au in such systems over timescales of 30–40 Myr. Consequently, the terrestrial-planet-forming region at < 2 au and the asteroid belt (> 2 au) often remained massive in these systems. As such disk conditions cannot form Mars and the asteroid belt, and late instabilities are currently disfavoured (Methods), these results imply that Jupiter and Saturn did not experience long-term deep locking in the 2:1 MMR during terrestrial-planet formation if an extended massive disk existed beyond ~1 au after gas dispersal. Instead, the success of our scenario based on JSCE, the evidence favouring an early instability and the existence of ~Mars-Earth-mass bodies during giant-planet formation indicate that Jupiter and Saturn experienced a near-resonance interaction in the 2:1 MMR.

*3.3 Jupiter–Saturn residual migration.* To test the effects of residual migration, we considered systems in which Jupiter and Saturn started on orbits such that their orbital period ratio PS/PJ was initially in the 2.3–2.45 range. These orbital configurations were set to mimic the orbital states of the giant planets acquired immediately after their instability. We also placed the four terrestrial planets on dynamically cold orbits ($e_0 = 0.001$ and $i_0 = 0.1°$) at their current semimajor axes. Then, following fictitious forces as implemented in the MERCURY code, Jupiter and Saturn smoothly migrated to their current orbits following an exponential behaviour with exponent (-t/tau); two values of tau were tested: 1 and 10 Myr. A total integration time of five times tau was used, resulting in 5- and 50-Myr timescales for the residual migration simulations described here. Overall, the results indicate that the orbit of Mercury could have been excited to eccentricities of $e \sim 0.1$–$0.3$ and inclinations of $i \sim 7$–$10°$ during this residual migration. Therefore, in our scenario framework, after Mercury's formation in the inner regions of the disk, Mercury's current orbit could be explained by further excitation caused by the residual migration of Jupiter and Saturn. We also tested the effects of residual migration on hypothetical asteroids concentrated at 2.2-3.2 au and possessing a wide range of eccentricities (0-0.2) and inclinations (0-38°). The final orbital structures were similar to the initial one, suggesting that the effects of residual migration were subtle for such an asteroid belt. A more detailed investigation of residual migration in the inner solar system is left to future work.

*3.4 Giant planet instability.* First, we considered 5-giant planet systems containing fully formed Jupiter, Saturn, and three icy giants. By employing fictitious forces to mimic the gas-driven migration of giant planets during the early solar system (a common technique used in past studies[51,55,57]), the five giant planets acquired mutual resonant orbits described by 2:1, 4:3, 3:2, 3:2 MMRs. This multi-resonant configuration is one of the preferred in past studies[45,57]. After slightly varying the initial conditions for these simulations, we obtained 357 unique resonant systems. In the first stage of our instability simulations (inst-S1), consistent with our current understanding of the outer solar system[24], we added a 20 ME dynamically cold trans-planetary disk at 20-30 au to each system (the disk contained 4000 massive objects). Then, by randomizing the mean anomaly of the icy giants to force early instabilities, we allowed the systems to evolve until the onset of the instability. Next, to increase the likelihood of obtaining "good" instability outcomes, we selected 49 post-instability systems that resulted in Jupiter, Saturn, and two icy giants and PS/PJ close to 2.5 within 20 Myr. Then, we inspected the orbital output of the selected systems and identified the approximate timing of imminent instability ($t_{inst}$). In the next stage, we used the orbital state of the giant planets and disk objects of each system at $t_{inst}$ as the initial conditions of new instability simulations (inst-S2). In addition, we also added 5271 asteroids beyond 2 au to each of these 49 seed systems. These asteroids were obtained at the end of JSCE based on the standard disk (Methods). To increase number statistics, we created additional 231 systems based on the seed systems by varying the mean anomaly of the asteroids, thus totalling 280 systems in inst-S2 simulations. These simulations evolved for 10 Myr. As expected, most of these systems experienced instabilities within a few Myr. Next, we selected post-instability systems of inst-S2 that experienced PS/PJ = 2.3-2.5 and Jupiter's mean eccentricity larger than ~0.03 (sample A: 41 systems) or ~0.04 (sample B: 29 systems) during the instability (note: sample A contains sample B plus 12 systems). These optimal instability systems are roughly consistent with the best outcomes found in similar works[51,57]. The instability was deemed finished when PS/PJ was the closest to 2.49 for each system ($t_{end}$). Conversely, when a system reached a time equals $t_{end}$, we assumed that the giant planets acquired their final orbits. At this point, we replaced the giant planets with the real ones on their current orbits and considered the orbital state of the asteroids found at $t_{end}$ as the initial conditions of long-term simulations for each system (4.1 Gyr, for consistency with our main simulations). A control system without instability consisting of the four giant planets and the 5271 asteroids was also evolved for 4.1 Gyr. Finally, we analyzed the orbital structure and dynamical depletion levels of the obtained asteroid belts in the control and the 41 optimal instability systems. Compared to the control system, we found that the depletion of asteroids increased by 2-78% (median 27%) for sample A, whilst we obtained 12-78% (median 34%) for sample B.

*3.5 Generating a near-2:1 MMR Jupiter–Saturn.* Similarly to the methods of ref.[25], we added Mars-Earth mass bodies to our multiple giant-planet systems described above (Section 3.4). We assumed various initial orbits for these bodies according to the location of internal/external MMRs with Jupiter ($a_{0,J}$ ~ 5.6 au) or Saturn ($a_{0,S}$ ~ 8.9 au). This procedure typically resulted in $a_0$ = 7.3 au, 10.3 au, and 10.8 au as the initial orbits of the planetary bodies. The masses tested for the planetary bodies were 0.1 ME, 0.2 ME, 0.5 ME, 1 ME, and 5 ME. No planetesimals were present in the disk. In the first batch of simulations, we considered solely the Jupiter–Saturn pair locked in 2:1 MMR and added a single planetary body. In the second batch of simulations, we considered our 5-giant planet multi-resonant system with and without an additional planetary body. Finally, our resonant Jupiter–Saturn pairs exhibited slightly different initial eccentricities across the various systems explored (~0.005 to 0.05). Considering that these simulations were highly stochastic (most experienced instabilities), we performed hundreds of simulations to probe some possible outcomes. Nevertheless, this investigation was not exhaustive, so dedicated studies are warranted to better understand the chaotic evolution of Jupiter and Saturn before the instability.

4. JUPITER–SATURN IN THE NEAR-2:1 MEAN MOTION RESONANCE

In a chaotic system, minimal differences in the initial conditions can yield distinct outcomes often described by exponential diverging behaviour[150]. For example, the orbital evolution of the planets in the solar system is chaotic, despite its long-term stability over 4.5 Gyr (ref.[151,152,153]). Here, although the initial conditions of Jupiter and Saturn were similar and both planets could remain stable during their near-resonant interactions in the 2:1 MMR over Myr, various frequencies associated with their orbital elements, such as the secular frequency associated with the longitude of the perihelion, wandered over wide ranges of values, clearly indicating chaotic behaviour (Figure S2). The orbital evolution experienced by our near-2:1 MMR Jupiter–Saturn pairs was similar to the findings of ref.[25]. Indeed, both planets exhibited similar libration/circulation of the two associated resonant arguments as defined in that work[25]. We think this chaotic behaviour arises when Jupiter and Saturn evolve with motion near the boundary of resonant motion and slow circulation in the 2:1 MMR (ref.[60]). Such an orbital configuration has also allowed the chaotic capture of Jovian Trojans during planetary migration[44,154]. Finally, although Jupiter and Saturn behave chaotically before the instability, the effects generated in the disk beyond ~1 au and the post-instability system outcomes were statistically similar among the various simulation runs.

An additional reason for our initial conditions is that the Jupiter–Saturn chaotic excitation (JSCE) mechanism seems much stronger when Jupiter and Saturn evolve near their 2:1 MMR rather than the 3:2 or other MMRs[25]. Therefore, as disk gas dynamics usually result in both planets experiencing the 3:2 or 2:1 MMR, our results suggest that Jupiter and Saturn evolved towards the latter. Perhaps the gas disk was of low mass to allow the planets to park near the 2:1 MMR instead of passing through it toward other internal MMRs, such as 3:2 (ref.[52,53,54]).

5. ADDITIONAL RESULTS AND IMPLICATIONS

*5.1 Global properties of terrestrial planet systems.* When analysing system properties, we found that the majority (60–80%) of our analogue systems satisfied the < 2AMD constraint depending on the disk model (Table S5). In particular, representative disk Ix yielded the highest successful fractions. Briefly, most of our systems remained dynamically cold by the end of the terrestrial-planet formation. This result is encouraging because including fragmentation or considering a higher number of planetesimals in the model would improve the system AMDs via enhanced dynamical friction[33,89]. Furthermore, approximately 85–95% of all systems satisfied the 0.5–2RMC constraint. Similar to previous models, our system RMCs were concentrated below the solar system value with medians RMC = 52–64. Nevertheless, these systems produced better RMC distributions than those found for systems containing the Venus–Earth pair in the early instability scenario[6]. However, because the RMC is highly sensitive to both the orbits and masses of Mercury and Mars, further improving RMCs will require detailed investigations of disk properties to increase the odds of producing optimal 4-P systems closely resembling the inner solar system. In fact, the AMD and RMC alone should not be used as primary success criteria for terrestrial system classification[72]. Instead, these metrics offer only complementary insights about the system's properties. Finally, most of the mass in embryos likely concentrated within 1.5 au in the disk, which facilitated the formation of compact analogue systems. Indeed, no individual planets were found within the asteroid belt in our 221 analogue systems.

*5.2 Mercury formation.* Out of 122 Mercury analogues obtained in our model (Table S4), 47 were found in 4-P systems and 75 formed in 3-P systems, which simultaneously contain analogues of the Venus–Earth pair and Mercury or Mars. Approximately 75% of these systems contained either a Mars-like planet (4–5 times Mars mass on average) or one with too small a mass (< 0.5 times Mars mass). Therefore, most of these 3-P systems can be considered almost as good as 4-P systems. Overall, including extended inner regions in the disk (disk Ix) yielded systems with better

Mercury–Venus orbital separations (real one $a_V$-$a_M$ = dMV = 0.34 au). Specifically, 30% of these Mercury–Venus pairs fell within the success range 0.67–1.33 × dMV (vs ~10% for the other disks combined). Similar results were obtained when considering the criterion 0.67–1.33 × orbital period ratio of both planets, PV/PM: 45% (vs 15%). However, these relatively low fractions indicate that model improvements are warranted. Despite the difficulty in satisfying this constraint[19,21,96,144], these results suggest that an inner region component with slightly distinct boundaries and total masses could explain the observed separation of both planets.

Although our simulations produced individual analogue systems containing Mercury analogues with masses comparable to Mercury (~0.025–0.10 ME), our planet analogues' median masses were generally about two to three times the mass of Mercury. First, we think further exploring the initial model conditions may reveal inner-region properties capable of producing more Mercury analogues with masses comparable to the real planet. In particular, confirming investigations reported in the literature[19,21,143], we found that the mass distribution and total mass in the inner region seemed to play essential roles in determining the final mass of Mercury analogues. Second, including fragmentation in the simulations may also help produce less massive Mercury analogues. Combining these two factors could lead to successful results concerning Mercury's small mass (i.e., median masses close to 0.055 ME).

Disk Ix successfully formed several Mercury analogues because the innermost embryo (protomercury) was typically located inside 0.5 au at the start of the simulations. Later, the analogues experienced modest variations in the semimajor axis due to collisions or gravitational scattering with other embryos within the inner region. Our Mercury analogues acquired their final masses by continuously accreting local and distant objects in the disk. Planetesimals dominated this accretion, so a substantial flux impacted the Mercury analogues during their formation. Among these planetesimals, a fraction possessed large eccentricities and inclinations that led to highly energetic collisions (e.g., Figure S4a,b). Because JSCE excited the disk beyond ~1-1.5 au in only a few Myr, presumably objects in the Mercury-forming region also experienced a similar bombardment of planetesimals. Therefore, these results support the hypothesis that Mercury's local building blocks and the forming Mercury were enriched in iron by cratering erosion[95,155] (i.e., resulting in Fe core fractions greater than the canonical 30%). It is also possible that other iron enrichment processes operated in the Mercury-forming region[21]. In conclusion, Mercury's iron core could result from the accretion of iron-rich building blocks and surface/mantle erosion experienced by the planet during its formation. Alternatively, hydrodynamical simulations suggest mantle-stripping collisions with one or a few embryos could explain Mercury's iron core[110,156]. Nevertheless, terrestrial-planet formation models testing this hypothesis could not explain Mercury's core fraction (~75%) and orbit/mass consistently[21,91,96]. On the other hand, these models explored only a limited range of initial conditions.

In summary, a dedicated study considering our favoured protoplanetary disk, giant-planet conditions, and the inclusion of erosion processes or fragmentation in N-body simulations would be an exciting avenue to discriminate the best Mercury-formation scenario. In addition, such studies would also allow us to understand better why Mercury seems peculiar compared to the other terrestrial planets. However, as simulations involving Mercury formation are very computationally intensive, we leave these more detailed investigations for future work.

*5.3 Venus and Earth formation.* We obtained a large number of Venus–Earth pairs (221) belonging to 3-P/4-P systems obtained from all simulations in this study (Table S4). As discussed in the main text, disks with initial mass concentrations within the core region (disks B, C and Ix) more often matched the mutual distance between Venus and Earth with up to 70% success (disk Ix). However, even for disks without such concentrations, the success rates were approximately 20% (disk D), 30% (disk A) and 40% (disk E) (Table S1). Intriguingly, these findings suggest that the small mutual distance of the Venus–Earth pair resulted from a peculiar initial mass distribution around ~0.8-1 au rather than the effects of the JSCE mechanism.

Both Venus and Earth analogues typically accreted 99% (70%) of their final masses from objects that were initially within ~2 au (0.8–1.2 au) in the disk. Clearly, the core region and adjacent subregions were the primary sources of building blocks for these planets. Nevertheless, the feeding zones of the planets were slightly shifted inward and outward for Venus and Earth, respectively, in agreement with past work[19,139,157].

*5.4 Late accretion of Earth and the other terrestrial planets.* Because JSCE quickly stirred asteroids beyond 2 au, there was a preference for remnant objects within 2 au to accrete late (> 100 Myr) after Earth's formation. As these objects originally formed at ~0.8–1.8 au in the disk, Earth accreted mostly dry enstatite and ordinary chondrite (EC and OC) objects during its late accretion. Also, as our disks consisted of 10–20% C-asteroids at < 2 au, this would explain the small contribution of carbon chondrite (CC) materials during the late stages of Earth accretion[38,158]. Conversely, Mercury, Venus, and Mars probably experienced similar late EC/OC accretion histories. Recent models of Mercury formation and Venus' coupled atmosphere–internal evolution are consistent with this picture[120,144]. In addition, the captured trans-Jovian asteroids summed to a total mass comparable to the current asteroid belt and concentrated beyond 2.5 au (ref.[43]), so their contribution to late accretion was negligible.

*5.5 Bulk compositions of Earth and Mars.* The composition of the terrestrial planets can be inferred from mixing models of elemental and isotopic compositions[105,140,158,159,160,161,162,163,164]. However, the estimates of the bulk compositions of Earth and Mars have considerable uncertainties. Another question is whether Earth accreted some of its building blocks from a disk component for which no representative meteorites have been identified in the current collections[160,161,165]. In short, there is still no consensus on the best model regarding the bulk compositions of terrestrial planets. Nevertheless, we consider the chondritic models below as proof of concept that our planet analogues could reproduce the compositions of Earth and Mars. More detailed modelling is beyond the scope of this paper.

Despite the uncertainties, the mixing models agree that Earth is made mostly of enstatite chondritic materials. For instance, ref.[159] suggested that Earth comprises 59-85%, 12-34% and 2-9% enstatite, ordinary and carbonaceous chondritic materials, respectively. For Mars, the intervals are roughly <68%, >32% and <3%, respectively. Here, to estimate the bulk composition of our Earth and Mars analogues, consistent with our primordial asteroid belt taxonomy (Methods), we considered three simple chondritic models with the following fractions: i) EC100% (0–1 au), OC80%-CC20% (1–2 au); ii) EC100% (0–1 au), EC20%-OC70%-CC10% (1–1.5 au), OC90%-CC10% (1.5–2 au) and; iii) EC90%-OC10% (0–1.2 au), OC90%-CC10% (1.2–2 au). Only the planet analogues formed in 4-P systems were considered in this investigation. We found that Earth analogues acquired on average 61/67/70% EC, 30/28/27% OC and 8/4/3% CC materials, whilst for Mars analogues, these results were 55/62/65% EC, 35/33/32% OC and 9/5/3% CC materials for models i/ii/iii, respectively. Although these chondritic models are not unique and similar models could lead to successful results, these results are nevertheless compatible with the literature cited above and within the involved uncertainties. Finally, our asteroid belt model suggests that the disk component at <0.8 au (roughly the inner region) did not provide objects that could survive as asteroids today. Thus, this might be the region of the protoplanetary disk that provided the "missing materials" to Earth and the other planets[140,165].

*5.6 Mars formation.* 146 Mars analogues formed in our analogue systems with orbits and masses closely akin to the actual planet. In particular, 99 and 47 Mars analogues formed in 3-P and 4-P systems, respectively (Table S4). Overall, the little dependence of the results on model details indicates that the JSCE scenario is robust in reproducing Mars. In addition, because 40–50% of the final masses of our Mars analogues were fed by objects located beyond 1 au, JSCE's perturbations helped to prevent our disks from producing overmassed Mars analogues.

As discussed in Section 1, Mars could have formed in timescales slightly longer than the canonical 10 Myr if its formation was protracted[98,99,100,101,102]. Here, the success rates for the formation timescales of our Mars analogues were ~10-40% in the main disk models (Table S5). Similarly, the median formation timescales were marginally compatible with the upper limits of our success criteria. However, Mars analogues typically experienced 1-3 giant impacts during formation. In particular, a formation timescale longer than ~15 Myr was often determined by the time of the last giant impact in unsuccessful cases. Further studies are warranted to understand better the role of giant impacts in the accretion history of Mars. Also, aiming at obtaining shorter Mars' formation timescales, future N-body simulations could reveal more favourable initial conditions, especially focusing on embryo individual mass, orbital range of embryos in the disk, the total mass in the disk beyond ~1 au (and the ratio of that mass in embryos and planetesimals), and dynamical evolution of the Jupiter–Saturn pair.

Finally, our Mars analogues typically experienced the last collision by impactors of ≥ 0.005 and ≥ 0.001 ME masses after medians of 63–109 and 134–142 Myr, respectively. Such collisions are consistent with those proposed for the origin of the dichotomy and the Borealis basin on Mars[64,166].

*5.7 Water delivery to the four terrestrial planets.* By analysing the water mass acquired by our planet analogues formed in the standard disk (Methods), we found that typically 0.01%, 0.01–0.5%, 0.5–10% and 10–40% WMFs for regions at < 1.5 au, 1.5–2 au, 2–2.5 au and 2.5–3.5 au were required to satisfy the terrestrial planets' water constraint (Section 1). Alternatively, if the disk were relatively dry at < 2 au (< 0.01%), the > 2 au region would need to be more water-rich with WMFs > 20%. Overall, our results imply that: 1. The inner regions of the disk at < 2 au contained more water than previously thought; 2. The disk was several times more water-rich at > 2 au than at < 2 au. Recent observations and analyses of S-type asteroids, meteorites, and primitive organic matter support point 1 (ref.[38,39,167,168]), whilst measurements of WMFs of 10–20% in C-asteroids[123,169], the expectation of high WMFs in D/P-asteroids[170,171] and the contribution of trans-Jovian objects captured as asteroids beyond ~2–2.5 au (ref.[40,43,85,172]) support point 2. The water ice line probably contributed similarly by enhancing the formation of water-rich asteroids in that region[26,80,81,82]. Finally, the contamination of C- and D/P-asteroids in our composition model (Methods and Section 5.8) is consistent with the above results regarding higher WMFs in the protoplanetary disk.

Based solely on the successful WMF models (Table S2), the median WMFs acquired by Mercury analogues in the standard disk varied within the interval $7.5 \times 10^{-5}$–$9.4 \times 10^{-4}$ over these models. Similar values are expected in representative disk Ix because our Mercury analogues generally acquired their water from objects beyond ~1.5 au regardless of the disk considered. This result strengthens the case that Mercury accreted non-negligible amounts of water during its formation. By performing a similar analysis for our Venus, Earth and Mars analogues, we obtained the following median WMF intervals acquired by these planets: $4.2 \times 10^{-4}$–$1.1 \times 10^{-3}$, $5.0 \times 10^{-4}$–$1.2 \times 10^{-3}$ (~2–5 Earth's oceans) and $2.8 \times 10^{-4}$–$1.5 \times 10^{-3}$, respectively. Therefore, Venus, Earth and Mars probably acquired significant water masses during their formation. Similar to Earth, JSCE allowed a quick delivery of water-rich objects initially located beyond ~1.5 au to the other terrestrial planets.

The 80 WMF models considered in this study are not unique, and other combinations of individual water fractions in each disk region could lead to successful results. In particular, as suggested by our results, more water-rich disks not modelled here could potentially explain the WMFs of Venus, Earth and Mars, thus yielding Earth with higher median water contents and higher contribution from disk objects located within ~1.5 au. Finally, a caveat is that the water acquired by our obtained planets represents upper limits for the models considered because water retention was not 100% efficient as assumed in our calculations[173].

*5.8 Origin and evolution of the asteroid belt.*

*a.1) Formation: general picture.* Giant-planet migration/instability and terrestrial-planet formation played essential roles in dynamically shaping the asteroid belt[6,16,17,18,32,35,84]. The orbital architecture of the current giant planets predicts an asteroid belt stable up to ~4 au. However, asteroids concentrate highly within 3.25 au, without a substantial population of asteroids located beyond this region. This fact implies that the giant planets may have played essential roles in dynamically sculpting this distant region (~3.25–4 au) or that the primordial asteroid belt did not extend beyond 3.25 au during the infancy of the solar system. Indeed, there is evidence that distant asteroids located beyond 3.25 au originated as captured trans-Jovian objects[84,172].

Our scenario reproduced all of these features. An eccentric Jupiter gravitationally destabilised the primordial asteroid belt beyond ~3.2 au, and JSCE strongly shaped and dynamically depleted the primordial asteroid belt. Later, captured asteroids populated the outer regions of the belt at the end of planetary migration (ref.[43], this study). As discussed in the main text, we concluded that the admixing of different types of asteroids in the current asteroid belt is evidence of local asteroids that survived the various perturbations by the giant planets (mainly JSCE) and captured asteroids from trans-Jovian objects that acquired stable orbits after planetary migration.

*a.2) Formation: additional discussion and caveats*. The asteroid belt probably experienced four main stages of evolution during solar system history: 1. disk gas dispersal, 2. JSCE, 3. instability, and 4. long-term. When investigating asteroid belt formation, our main simulations took into account stages 2 and 4, whilst our auxiliary simulations considered the effects of stage 3, as discussed below.

Concerning stage 1, if Jupiter and Saturn acquired their current ($a_J$ = 5.20 au, $a_S$ = 9.55 au, $e_{JS}$ ~ 0.05) or more eccentric orbits ($a_J$, $a_S$ idem, $e_{JS}$ ~ 0.1) during the solar system's gas dispersal stage, the sweeping of the nu5 secular resonance could have depleted the primordial asteroid belt beyond ~2 au (ref.[94,141,142,174,175]). Nonetheless, given the various parameters associated with disk gas dynamics and disk-giant planet interactions, past studies have explored only a limited set of parameters and giant planet configurations regarding this mechanism. For instance, as predicted by hydrodynamic models (Methods), the cases of Jupiter and Saturn locked in a MMR (3:2 or 2:1) with low-moderate eccentricities (~0.01-0.1) remain unexplored. Other effects, such as gap-opening, also remain unexplored in recent studies[94,141,142]. In addition, if the gas dispersal were fast (< several hundred kyr) as considered in some models[40], then the effects of the sweeping nu5 secular resonance would probably be much less significant. Similarly, for rapid gas dispersal *after* Jupiter formation (~0.1 Myr timescales or shorter), these effects would be negligible[174]. Another critical point is that hydrodynamic modelling predicts many possible eccentricity evolutions for Jupiter and Saturn interacting in the 2:1 MMR, so it is possible that the giant planets experienced $e \leq 0.05$ during that interaction[56,57,174]. Finally, residual embryos/planetesimals may have reduced the eccentricities of the giant planets via dynamical friction during stage 1. Considering the reasoning above, we conclude that our assumption that the primordial asteroid belt was massive and extended to 3.5 au after disk gas dispersal is conceivable.

Regarding stage 3, the role of the instability in perturbing the primordial asteroid belt depends on the instability's details and the belt's initial dynamical state. For instance, Jupiter's orbital evolution behaviour (i.e., how "jumpy" the evolution is), duration of the instability, and scattering events with icy giants are important factors[6,16,17,84]. While the instability can stir an initially cold primordial asteroid belt, it can shuffle the eccentricities of an initially hot belt. Consistent with the latter scenario, for our excited primordial asteroid belt (by JSCE) (Figure S4a), there is evidence that the instability *does not* significantly alter the final orbital structure acquired by the asteroid belt after Gyr timescales[84,145]. Nevertheless, ref.[145] found that the instability increased long-term depletion in the asteroid belt by 60%. Based on our optimal instability systems, we found a median long-term depletion increase of 27–34% (Section 3.4). These simulations also confirmed that the asteroids that survived 4 Gyr after the instability have similar *a-e-i* structure compared to those in which the instability was simplified by instantaneously transporting the giant planets to their current orbits, as done in our baseline model and previous studies[45,71]. Despite that, we found that the instability increased the f6 ratio in our optimal systems. Compared to our control simulation in this investigation (f6 = 0.12), the optimal systems yielded f6 = 0.21–0.28, or an increase of 75-133%. Based on our main simulations, the representative asteroid belt yielded f6 = 0.16–0.19 after assuming reasonable proportions of local and captured asteroids (Methods). In this way, by considering the effects of instability and the final phase of residual migration[146] (Section 2.5), we obtained f6 = 1.75 × 0.4 × (0.16–0.19) = 0.11–0.13 or 2.33 × 0.4 × (0.16–0.19) = 0.15–0.18. Therefore, we estimate a final f6 is lying within 0.11–0.18 in this work.

Of course, our representative asteroid belt is not a perfect match for the real belt. For example, there is a deficit of asteroids at 2.1-2.3 au. This specific region contains 23/895 = 2.6% and 2/(159 or 318) = 0.6-1.3% of real and model asteroids, respectively. The latter fraction varies according to the proportion of local:captured asteroids adopted. Potential solutions to this discrepancy include asteroids' self-gravity[16] (not modelled here) and the instability[176]. Under the effect of these mechanisms, asteroids can experience radial displacements that could help populate the inner region within 2.3 au. Based on our auxiliary simulations of stage 3 described above, we noticed a small increase of asteroids in the inner region. Another potential solution is that slightly different initial conditions for the Jupiter–Saturn pair might reduce the depletion levels of primordial asteroids in the 2-2.5 au region.

Overall, as discussed in the main text, our asteroid belt model broadly explains the orbital structure of the asteroid belt.

*b) Small mass.* The current asteroid belt is up to several thousand times less massive than predicted at 2–4 au if the disk that formed the terrestrial planets followed the same mass distribution until 4 au (ref.[177]). As collisional grinding cannot explain this mass discrepancy, either the giant planets played significant roles in the mass depletion of the primordial asteroid belt or the latter was not massive[8,9,10,12,18,14,32,130].

Our model can potentially solve the puzzling small mass of the asteroid belt. At the end of terrestrial-planet formation (after ~100 Myr in our simulation framework; henceforth 't100'), the median masses and survival fractions for asteroid belts found in the standard disk were ~5.0 × 10$^{-3}$ ME and 0.70% (46 systems). After evolving these systems over 4 Gyr, 81% (445/551) of the asteroids were removed. These results were consistent with 4-Gyr depletion levels of 50–60% in belts based on the observed one[131,132], ~70% in post-Grand-Tack systems[45], and (50) 80% for systems that experienced (no) dynamical instabilities in ref.[145]. The initial orbital states of the asteroid belt that yielded 50–60% depletion in past work were similar to the observed belt, while our asteroid belts exhibited a wider range of eccentricities and inclinations as obtained directly from the simulations. Thus, our asteroid belts covered more unstable areas of orbital space resulting in higher removal fractions. Therefore, by combining the fractions above, we found a primordial asteroid belt depletion level of 99.87%.

In conclusion, the final mass of our representative asteroid belt resulted in $\sim 1 \times 10^{-3}$ ME in local asteroids. Considering the additional depletion caused by the instability of 27–34%, this estimate reduces to $\sim 7 \times 10^{-4}$ ME, which after combining with the complement mass in captured asteroids resulted in an asteroid belt with mass $(1.5\pm0.5) \times 10^{-3}$ ME depending on the fractions of local/captured asteroids adopted in the current asteroid belt. This result translates into a factor of 2-4 times the mass in the observed belt. We suggest the following hypotheses to reconcile this result with the observed mass of the asteroid belt: 1. The primordial asteroid belt was initially less massive (e.g., ≤0.5-1 ME); 2. The presence of small embryos in the primordial asteroid belt provided additional depletion[130,178]; 3. Given the spectrum of possible instability evolutions[9,22,51,57,58,67], the instability experienced by the giant planets yielded smaller fractions of local and captured asteroids.

We conclude that our representative asteroid belt acquired a small mass in reasonable agreement with observations.

*c) Composition taxonomy.* By applying our nominal model of primordial asteroid compositions to local and captured asteroids, we found that our representative asteroid belt could explain the currently observed orbital concentrations of S-, C- and D/P-type asteroids[15,123] (Figures 4, S7 and S8). We also tested several distinct fractions of S-, C- and D/P assigned to the local and captured asteroids to verify dependence on model parameters. Compared to the nominal model (S90~80%-C10~20% within < 2 au and S50%-C50% beyond > 2 au for local asteroids and C50%-DP50% for captured asteroids), a higher or lower fraction of C-asteroids beyond 2 au among local asteroids resulted in too many or too few final C-type asteroids within the same region. Conversely, increasing the proportion of S-asteroids (fewer C-asteroids) also led to too many final S-type asteroids in this region. Therefore, while a S50%-C50% contribution beyond > 2 au was favoured, a contribution of C-asteroids at < 2 au (i.e., fewer S-asteroids) was needed to match the final S-type asteroid distribution better. Although we could not strongly constrain the proportion of C-asteroids at < 2 au, matching the observed C- and S-type asteroids in the inner asteroid belt required C-asteroids with ~10–20% abundances within this region. About captured asteroids, a higher fraction of C-asteroids (smaller fraction of D/P-asteroids) resulted in too many final C-type and a lack of D/P-type asteroids beyond ~2.9 au. Conversely, a higher fraction of D/P-asteroids resulted in too many D/P-type and a paucity of C-type asteroids within the same region. Furthermore, the reproduction of distant S-type asteroids beyond 2.9 au may require a small contribution of S-asteroids (~5%) among the captured population. All these trends are valid for an asteroid belt consisting of comparable local and captured asteroids with proportions within a factor of 2.

For completeness, we created two additional WMF models based on the nominal composition model by assuming that the WMFs of C- and D/P-asteroids were 5% or 10% and 20% or 30%, respectively (models 27 and 28 in Table S2). The wetter WMF model of C-asteroids (10%) was successful for our Venus, Earth and Mars analogues, independent of the choice of WMF for D/P-asteroids. Therefore, we confirmed that our results were self-consistent regarding asteroidal compositions and water delivery.

*d) Late accretion and the inner solar system's bombardment.* After t100 and the Moon-forming giant impact, our results imply that Earth experienced non-negligible accretion by impactors over 4 Gyr. In the *a-e-i* element region (0.7–2 au, 0–0.65, 5–50°) that sourced 137 of the 142 impactors to Earth over 4 Gyr, there were initially 1985 particles and a total of 0.026 ME. Thus, 0.026 × (137/1985) = 0.0018 ME of mass accreted to Earth during this period. We also identified several late collisions with Earth well after t100 over timescales of hundreds of Myr, thus supporting the hypothesis that terrestrial-planet formation tail-end population caused a bombardment in the inner solar system[179,180]. These results were also consistent with the findings of ref.[46,64].

*e) The residual EE-belt.* In agreement with previous studies, our results also support the existence of a primordial disk component located beyond the orbit of Mars akin to the hypothesised 'E-belt' (ref.[71]). However, while the latter was assumed to be a relatively stable component of asteroids at $a = 1.6$–2.1 au, perihelia $q < 1.6$ au and $i < 20°$, our results instead predict an excited E-belt (EE-belt) with a more comprehensive range of eccentricities and inclinations persisting during and after terrestrial-planet formation. For $a = 1.6$–2.1 au and without restricting $q$ or $i$, we obtained a median of 0.015 ME for the EE-belt or ~25 times the mass of the current asteroid belt at t100. Approximately one order of magnitude less mass is obtained if $q < 1.6$ au and $i < 20°$ are applied. Briefly, our study revealed that an EE-belt could have arisen naturally during/after the formation of the terrestrial planets.

*f) Emptied primordial asteroid belt.* Notably, a long-term JSCE may have depleted the primordial asteroid belt entirely. In such a scenario, the asteroid belt might have formed due to the implantation of S-type objects scattered out from the terrestrial region during terrestrial-planet formation and implantation of trans-Jovian C-type objects during Jupiter–Saturn interactions and post-instability evolution[18,40,43]. However, it is currently unclear whether this alternative scenario could consistently explain the formation of the four terrestrial planets and the orbital structure, the f6 constraint, the small total mass and the taxonomy of the asteroid belt. See also Section 2 for complementary discussions.

## 6. MAIN FINDINGS AND IMPLICATIONS

Overall, after considering several plausible initial conditions for the early solar system in the framework of the JSCE scenario, our results identified those with good prospects for explaining the terrestrial planets and the asteroid belt. We scrutinised our terrestrial planet systems with unprecedented detail and identified several that successfully satisfied fundamental constraints in the inner solar system. These results imply that the terrestrial planets probably formed from a protoplanetary disk containing mass concentrated within a core region surrounded by a low-mass inner region and a mass-depleted and dynamically perturbed outer region. Furthermore, JSCE played a crucial role in shaping the disk. Finally, the asteroid belt formed by local asteroid remnants and captured trans-Jovian objects that survived until this period.

If correct, our main findings can be summarised as follow.

Our disks can produce 4-P systems and reasonably reproduce the orbits, masses, accretion histories and other aspects of the four terrestrial planets. In particular:

- Mercury formed within the low-mass extended inner region;
- The Venus–Earth pair formed from a mass-concentrated thin annulus within the core region;
- Mars formed within the mass-depleted outer region;
- The existence of the inner and outer regions allowed the formation of Mercury and Mars at the correct distances from Venus and Earth, respectively.

Earth experienced the Moon-forming giant impact(s) within 60 Myr after solar-system formation. The impactor(s) was(were) less massive than Mars, with a preference of $0.02 \leq m < 0.1$ ME. After the last giant impact, late accretion of dry EC/OC-like objects (originally formed at 0.8–1.8 au) contributed to < 0.01 ME of Earth's final mass.

Disks containing a drier region within 1.5 au and a wetter region beyond ~1.5-2 au yielded good prospects for delivering water to Earth and the other terrestrial planets. Earth's bulk water was acquired during the first 10–20 Myr of its accretion, before the formation of the Moon. The other terrestrial planets also acquired significant amounts of water.

Our representative asteroid belt can reproduce the $a$-$e$-$i$ orbital structure, the mixed populations of S-, C- and D/P-type asteroids, the small total mass and other features of the asteroid belt. Furthermore, at the end of the terrestrial-planet formation, a reservoir of primordial asteroids with $a = 1$–$2$ au and $i = 15°$–$50°$ sourced late impacts on Earth over several hundred Myr. Thus, terrestrial-planet formation tail-end asteroids can explain the inner solar system's bombardment.

Jupiter and Saturn interacted near their mutual 2:1 MMR for ~5–10 Myr timescales before experiencing their dynamical instability/migration and acquiring their current orbits. Under this orbital configuration, secular resonances associated with both the giant planets behaved chaotically, resulting in strong dynamical excitation and depletion of the disk beyond ~1–1.5 au, including >99% depletion of a massive primordial asteroid belt. This event allowed the formation of a small-mass Mars on a moderately excited orbit and a very low-mass and stirred asteroid belt. Then, after planetary migration was complete, captured asteroids from the trans-Jovian region populated the asteroid belt. Nowadays, the asteroid belt consists of comparable populations of local and captured asteroids.

**Supplementary Figures**

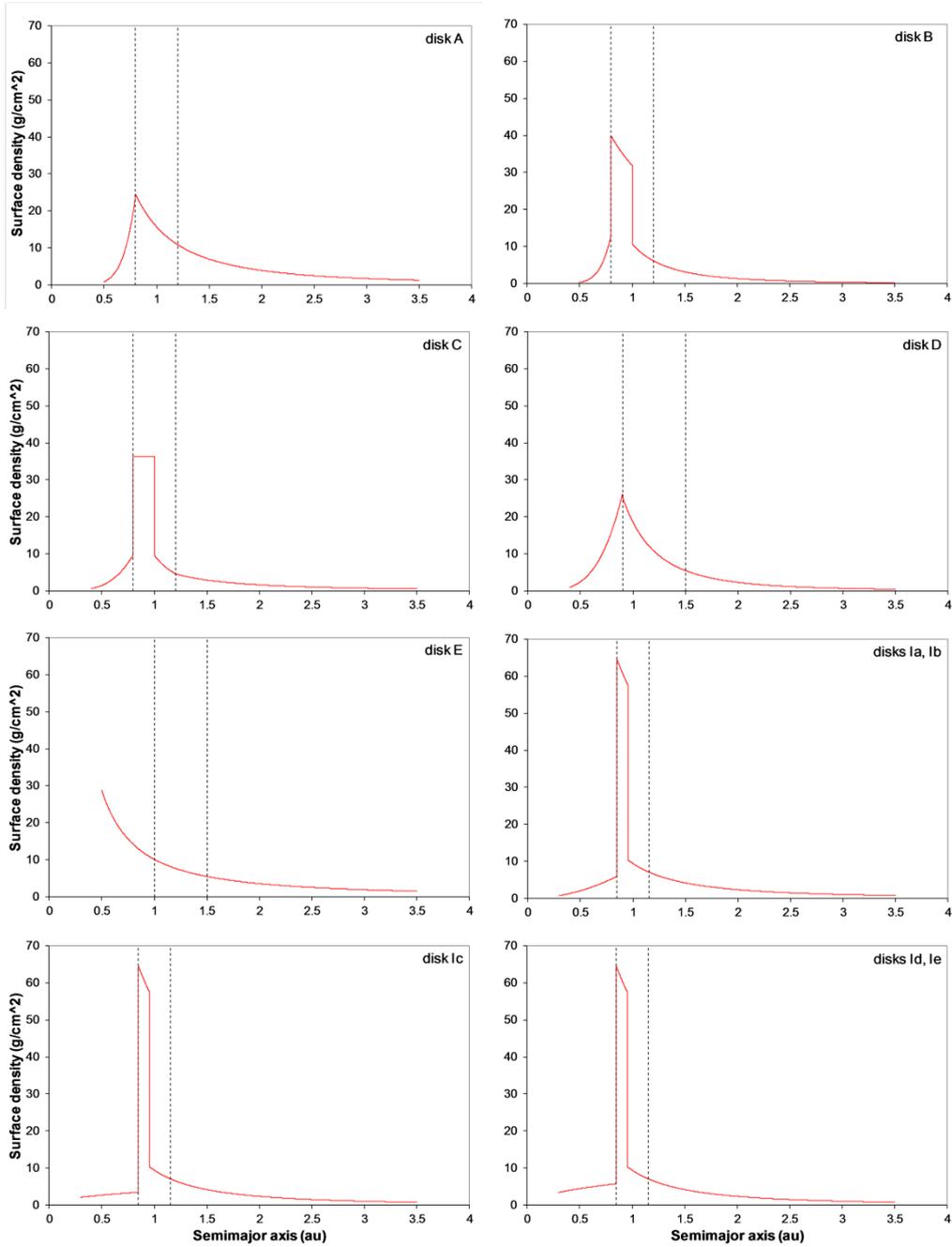

**Figure S1**. Initial mass distributions of the disk models used in this work. The dashed lines indicate the boundaries of the core region. See Methods and Table S1 for more details.

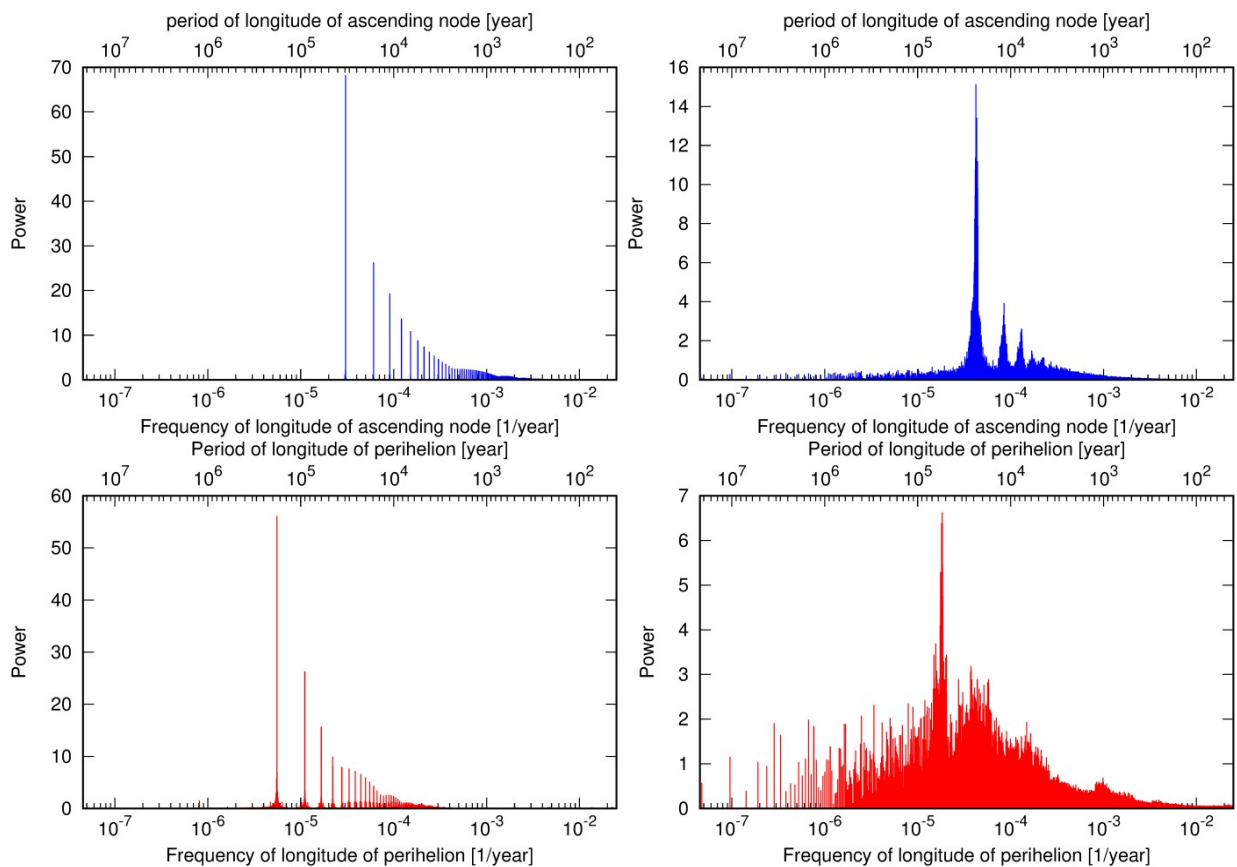

**Figure S2.** Power spectra obtained by Fourier analysis of the longitude of the ascending node (top panels) and longitude of the perihelion (bottom panels) of Saturn. The left and right panels illustrate Saturn's non-chaotic and chaotic behaviour from two simulations where the Jupiter–Saturn pair and a disk of primordial asteroids evolved to 10 Myr, respectively. In the chaotic case, Jupiter and Saturn were placed near their mutual 2:1 MMR.

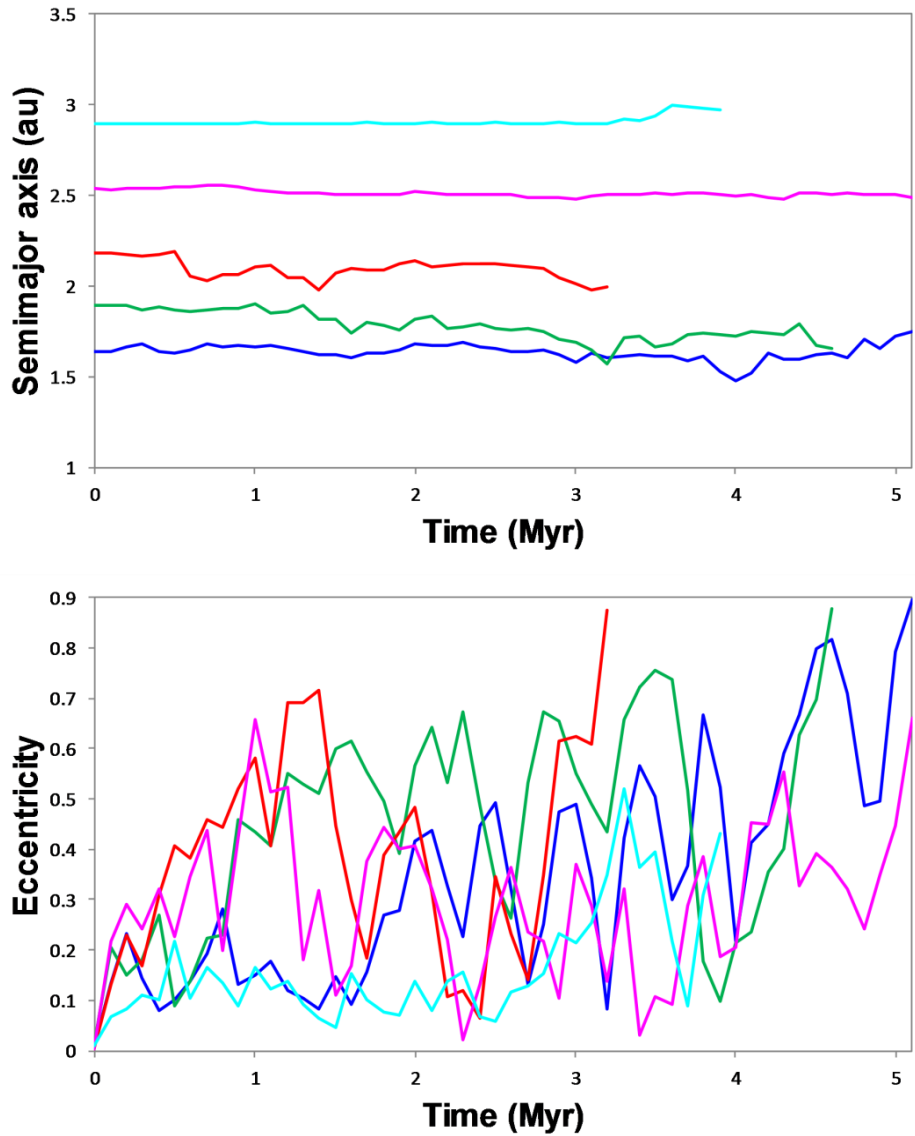

**Figure S3.** Orbital evolution of five selected asteroids of system #1 (Figure 2) that were lost from the solar system due to chaotic excitation induced by a near 2:1 MMR Jupiter–Saturn before the giant planet instability/migration. The asteroids also gravitationally interacted with the embryos and planets in formation, as indicated by changes in the semimajor axis.

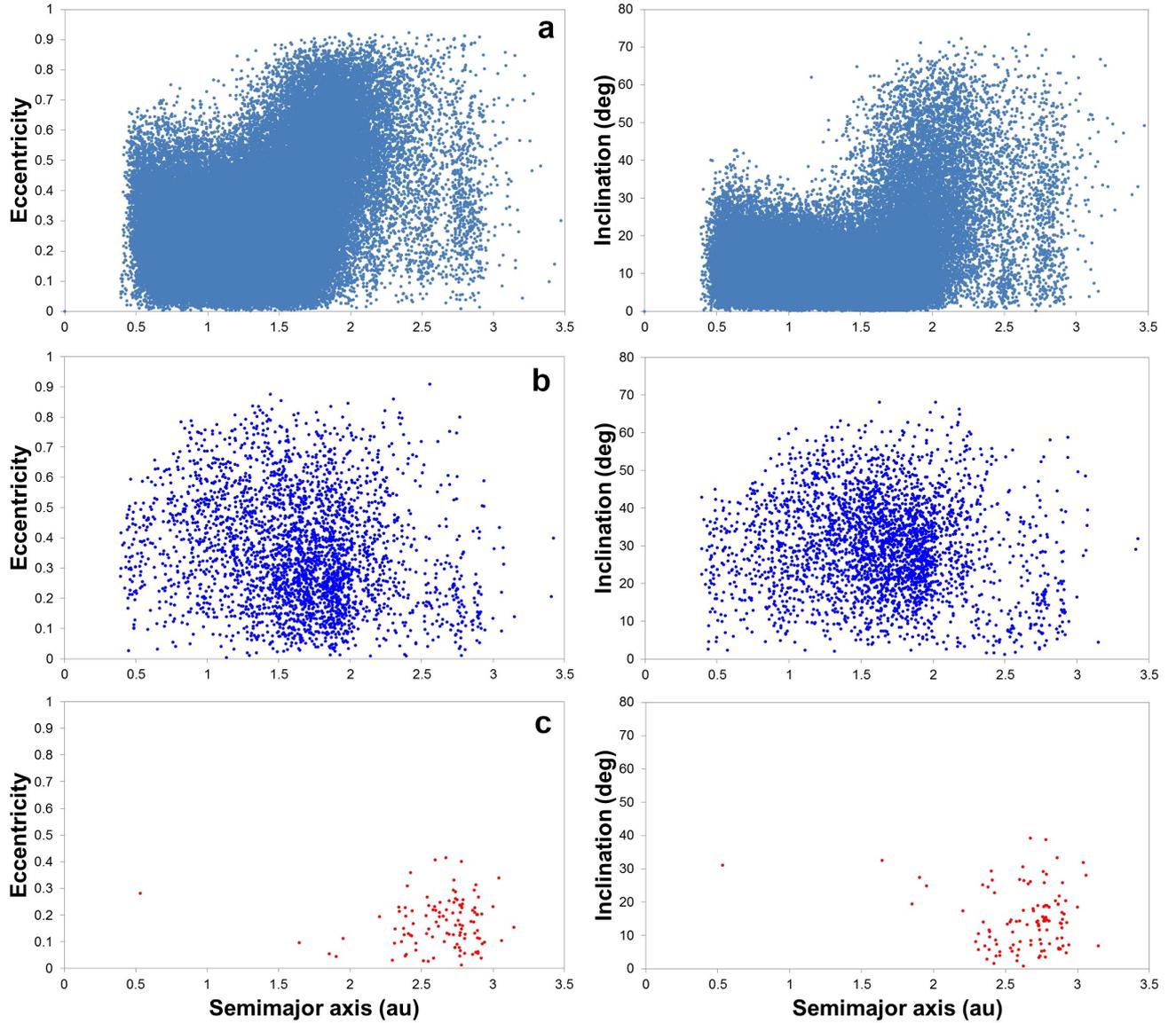

**Figure S4.** Composite of the orbital evolution of embryos and planetesimals initially placed on dynamically cold orbits at ~0.5–3.5 au in our standard protoplanetary disk (46 systems combined from disks A, B and C). **a)** After 5–20 Myr of chaotic excitation generated by near-2:1 MMR Jupiter–Saturn, the disk was strongly perturbed beyond ~1–1.5 au. **b)** After a further 100 Myr of post-instability evolution with the giant planets on their current orbits. **c)** After a further 4 Gyr of dynamical evolution with all the planets on their current orbits. Only local asteroids are shown. See Methods and Supplementary Information for more details.

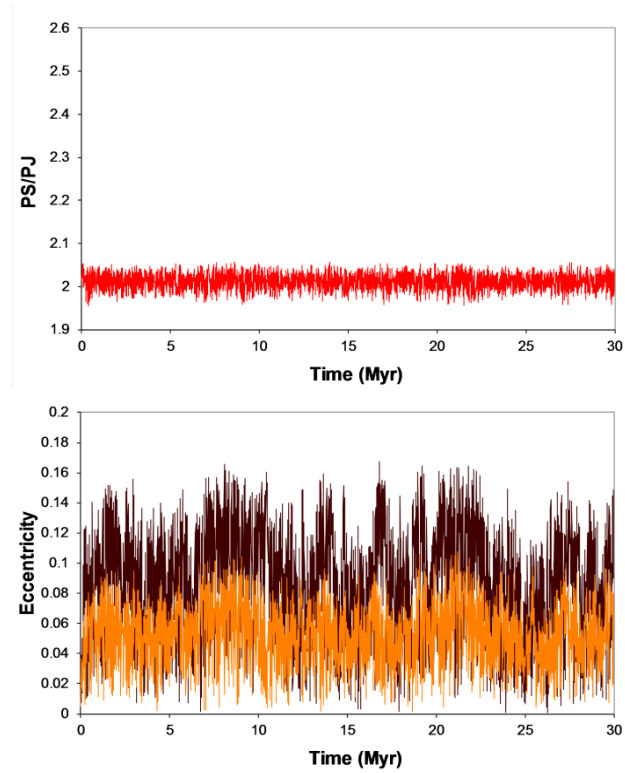

**Figure S5**. An example of how low-*e* Jupiter ($e_{0,J}$ = 0.005; orange curves) and Saturn ($e_{0,S}$ = 0.009; brown curves) locked in their mutual 2:1 MMR can evolve to a near-resonant configuration that generates chaotic behaviour for both planets. PS and PJ represent the orbital period of Saturn and Jupiter, respectively. Here, the simulation considered Jupiter, Saturn, and one 5 ME large body placed at 10.3 au outside Saturn's initial orbit ($a_{0,S}$ = 8.92 au). The object suffered gravitational scattering by Saturn at ~0.09 Myr and later diffused outwards until ejection from the system at 2.77 Myr. As a result, Jupiter and Saturn experienced chaotic orbital behaviour near the 2:1 MMR from 0.1 Myr until the end of the simulation at 50 Myr.

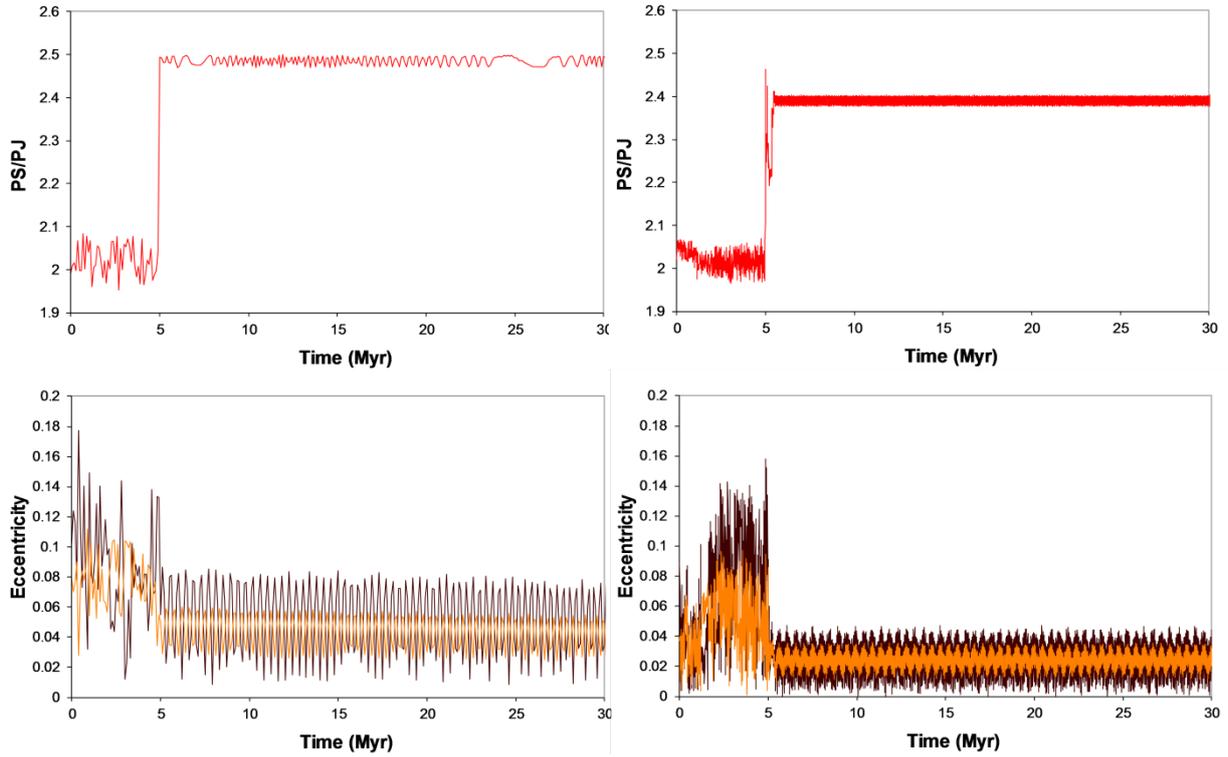

**Figure S6**. The typical orbital evolution of Jupiter (orange curves) and Saturn (brown curves) in our terrestrial planet formation model for an assumed instability timing of 5 Myr (left panels) and a self-consistent simulation containing Jupiter, Saturn ($e_{0,J} = e_{0,S} \sim 0.04$), and three icy giant planets initially locked in mutual MMRs (right panels). PS and PJ represent the orbital period of Saturn and Jupiter, respectively. In the latter simulation, the five giant planets were locked in a 2:1, 4:3, 3:2, 3:2 resonant chain. In both cases, the Jupiter–Saturn pair experiences a chaotic orbital behaviour near the 2:1 MMR during the first ~5 Myr of evolution, after which the giant planet instability occurs (spontaneously in the latter simulation). Also, the PS/PJ evolution is consistent regardless of the assumed initial instantaneous PS/PJ in both simulations.

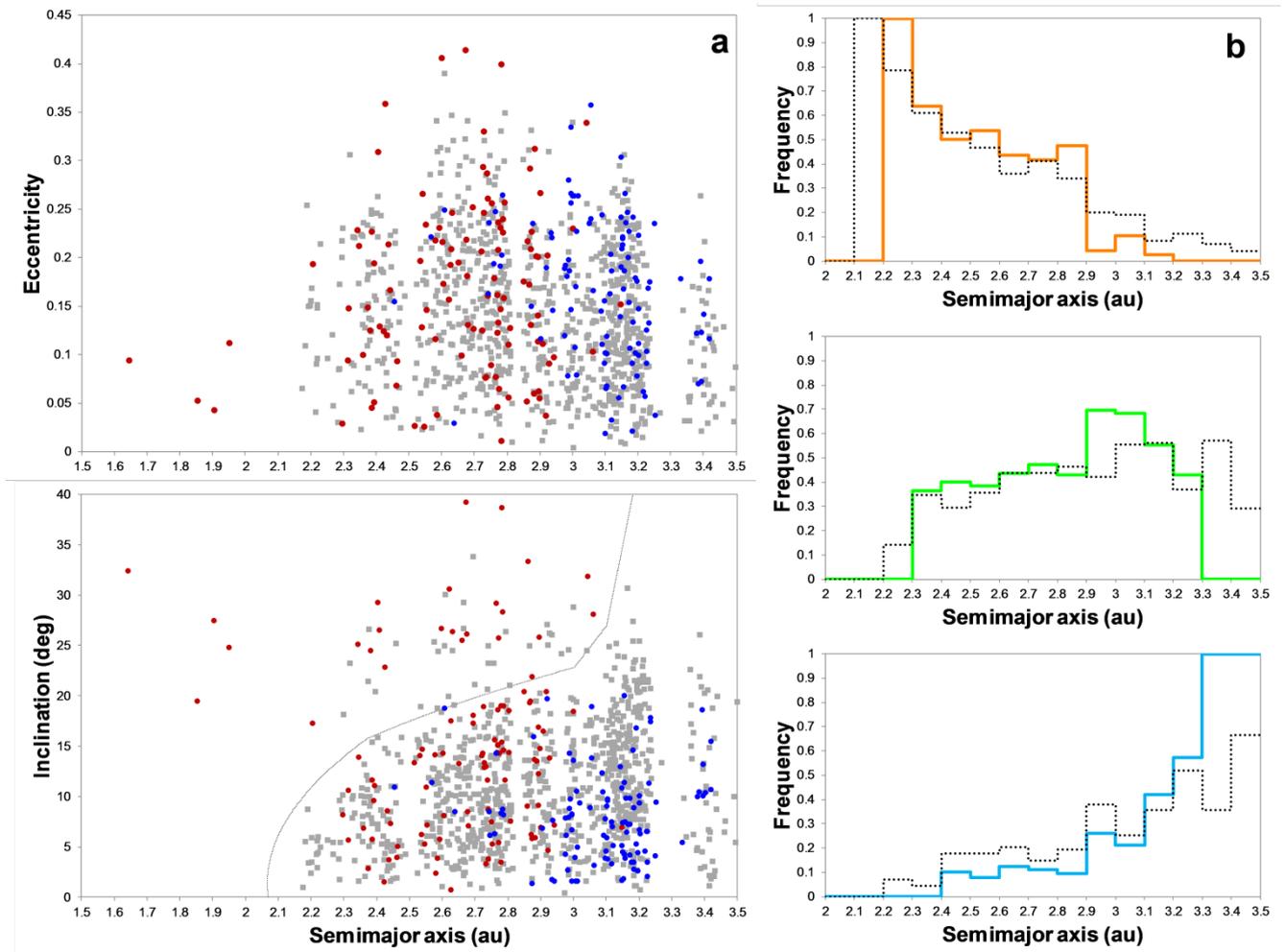

**Figure S7**. The same as Fig. 3, except that the proportion of local:captured asteroids is 50%:50%, and the composition gradient of captured asteroids is represented by C50%-DP50%.

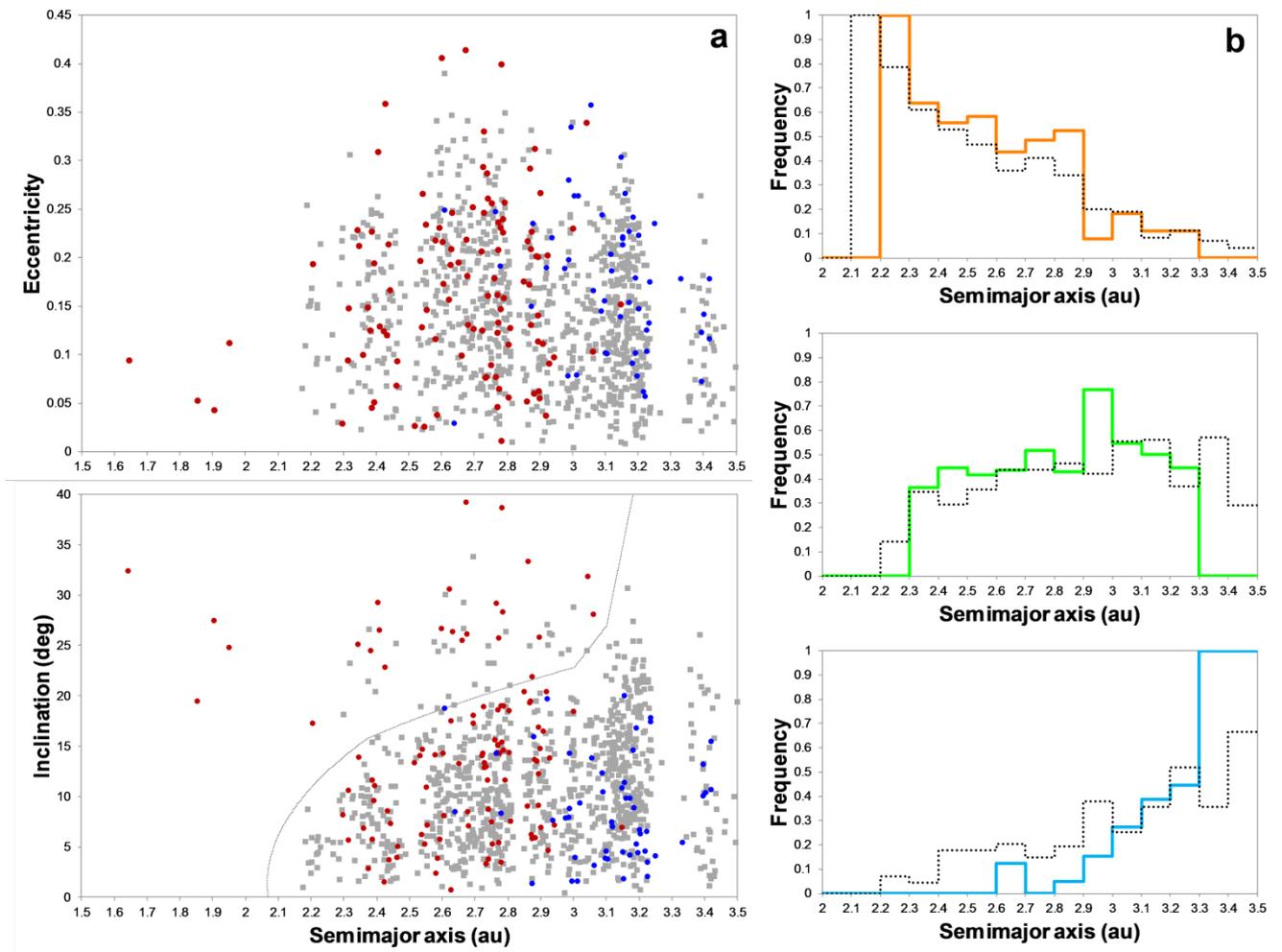

**Figure S8.** The same as Fig. 3, except that the proportion of local:captured asteroids is 67%:33%.

# Supplementary Tables

**Table S1.** Initial conditions in our simulations

| Disk | n. embryos [m_emb ($M_\oplus$)] | n. planetesimals (including asteroids) | n. of asteroids (2-3.5 au) | Total mass AB ($M_\oplus$) | Embryos a range (au) | $r$ in the inner and core regions | surface density slopes in the inner, core and outer regions | inner region (au) [total mass ($M_\oplus$)] | core region (au) [total mass ($M_\oplus$)] | outer region (au) [total mass ($M_\oplus$)] | instability time (Myr) | post-instability runs |
|---|---|---|---|---|---|---|---|---|---|---|---|---|
| A | 153 [0.005] | 6220 | 2068 | 2.07 | 0.5-1.2 | 2, 0.5 | 7, -2, -2 | 0.5-0.8 [0.40] | 0.8-1.2 [1.50] | 1.2-3.5 [3.96] | 10, 20 | 50, 50 |
| B | 155 [0.005] | 5441 | 1075 | 0.54 | 0.5-1.2 | 2, 0.5 | 8, -3(-1*), -3 | 0.5-0.8 [0.20] | 0.8-1.2 [1.92]* | 1.2-3.5 [1.37] | 10, 20 | 50, 50 |
| C | 105 [0.01] | 4380 | 1150 | 0.87 | 0.45-1.2 | 2, 1 | 4, -4(0*), -2 | 0.4-0.8 [0.24] | 0.8-1.2 [1.90]* | 1.2-3.5 [1.67] | 5-10 | 100 |
| D | 241 [0.005] | 5312 | 1232 | 0.94 | 0.4-1.5 | 2, 0.5 | 4, -3, -3 | 0.4-0.9 [0.85] | 0.9-1.5 [1.94] | 1.5-3.5 [1.67] | 10 | 50 |
| E | 21 [0.078] | 4738 | 2191 | 2.18 | 0.5-1.5 | 4, 1 | -1.5, -1.5, -1.5 | 0.5-1.0 [1.40] | 1.0-1.5 [1.07] | 1.5-3.5 [3.09] | 5-10 | 50 |
| Ia | 36 [0.02] | 4404 | 994 | 1.24 | 0.4-1.15 | 2, 0.5 | 2, -2(-1*), -2 | 0.3-0.85 [0.24] | 0.85-1.15 [1.70]* | 1.15-3.5 [2.47] | 5 | 50 |
| Ib | 35 [0.03] | 3767 | 994 | 1.24 | 0.4-1.15 | 4, 1 | 2, -2(-1*), -2 | 0.3-0.85 [0.26] | 0.85-1.15 [1.70]* | 1.15-3.5 [2.47] | 5 | 50 |
| Ic | 68 [0.01] | 4476 | 994 | 1.24 | 0.35-1.15 | 1, 0.5 | 0.5, -2(-1*), -2 | 0.3-0.85 [0.22] | 0.85-1.15 [1.70]* | 1.15-3.5 [2.47] | 5 | 50 |
| Id | 40 [0.02] | 4483 | 994 | 1.24 | 0.35-1.15 | 2, 0.5 | 0.5, -2(-1*), -2 | 0.3-0.85 [0.36] | 0.85-1.15 [1.70]* | 1.15-3.5 [2.47] | 5 | 50 |
| Ie | 38 [0.03] | 3811 | 994 | 1.24 | 0.35-1.15 | 4, 1 | 0.5, -2(-1*), -2 | 0.3-0.85 [0.37] | 0.85-1.15 [1.70]* | 1.15-3.5 [2.47] | 5 | 50 |

**Notes.** AB, asteroid belt; $r$, the ratio of total mass in embryos to that in planetesimals. We placed the embryos in the inner and core regions and planetesimals within all disk regions. Planetesimals located at 2–3.5 au represented the primordial asteroid belt. The embryos started with the same mass (m_emb) in a given disk model. The mass of each planetesimal was $5 \times 10^{-4}$ ME in the inner and core regions and within $5–10 \times 10^{-4}$ ME in the outer region. See Methods for other details. A summary of the initial disk conditions is illustrated in Figure S1.

* The following disks possessed mass concentration within the core region. Of the total mass in the core region, 1.50 ME was concentrated within 0.8–1.0 au for disk B (slope −1), 1.54 ME was concentrated within 0.8–1.0 au for disk C (slope 0), and ~1.28 ME was concentrated within 0.85–0.95 au for disks Ia–e (slope −1).

**Table S2.** Initial water mass fractions (WMFs) for embryos and planetesimals located initially in distinct regions of the protoplanetary disk and final WMFs acquired by planet analogues in the 3- and 4-terrestrial planet analogue systems

| # | <1.5 au | 1.5-2 au | 2-2.5 au | >2.5 au | >3 au | Mercury | Venus | Earth | Mars | Result |
|---|---------|----------|----------|---------|-------|---------|-------|-------|------|--------|
| 1 | $1.0\times10^{-5}$ | $1.0\times10^{-5}$ | $1.0\times10^{-3}$ | $5.0\times10^{-2}$ | | $1.2\times10^{-5}$ | $5.5\times10^{-5}$ | $6.9\times10^{-5}$ | $2.0\times10^{-5}$ | |
| 2 | $1.0\times10^{-5}$ | $1.0\times10^{-5}$ | $1.0\times10^{-3}$ | $1.0\times10^{-1}$ | | $1.2\times10^{-5}$ | $9.4\times10^{-5}$ | $1.2\times10^{-4}$ | $2.0\times10^{-5}$ | |
| 3 | $1.0\times10^{-5}$ | $1.0\times10^{-5}$ | $1.0\times10^{-3}$ | $3.0\times10^{-1}$ | | $1.2\times10^{-5}$ | $2.5\times10^{-4}$ | $3.1\times10^{-4}$ | $2.0\times10^{-5}$ | |
| 4 | $1.0\times10^{-5}$ | $1.0\times10^{-5}$ | $1.0\times10^{-2}$ | $5.0\times10^{-2}$ | | $3.2\times10^{-5}$ | $1.1\times10^{-4}$ | $1.2\times10^{-4}$ | $1.2\times10^{-4}$ | |
| 5 | $1.0\times10^{-5}$ | $1.0\times10^{-5}$ | $1.0\times10^{-2}$ | $1.0\times10^{-1}$ | | $3.2\times10^{-5}$ | $1.4\times10^{-4}$ | $1.7\times10^{-4}$ | $1.2\times10^{-4}$ | |
| 6 | $1.0\times10^{-5}$ | $1.0\times10^{-5}$ | $1.0\times10^{-2}$ | $3.0\times10^{-1}$ | | $3.2\times10^{-5}$ | $3.0\times10^{-4}$ | $3.9\times10^{-4}$ | $1.2\times10^{-4}$ | |
| 7 | $1.0\times10^{-5}$ | $1.0\times10^{-5}$ | $3.0\times10^{-2}$ | $1.0\times10^{-1}$ | | $7.5\times10^{-5}$ | $2.4\times10^{-4}$ | $2.7\times10^{-4}$ | $3.1\times10^{-4}$ | |
| 8 | $1.0\times10^{-5}$ | $1.0\times10^{-5}$ | $3.0\times10^{-2}$ | $2.0\times10^{-1}$ | | $7.5\times10^{-5}$ | $3.4\times10^{-4}$ | $3.9\times10^{-4}$ | $3.3\times10^{-4}$ | |
| **9** | $\mathbf{1.0\times10^{-5}}$ | $\mathbf{1.0\times10^{-5}}$ | $\mathbf{3.0\times10^{-2}}$ | $\mathbf{3.0\times10^{-1}}$ | | $\mathbf{7.5\times10^{-5}}$ | $\mathbf{4.0\times10^{-4}}$ | $\mathbf{5.0\times10^{-4}}$ | $\mathbf{3.3\times10^{-4}}$ | **OK** |
| **10** | $\mathbf{1.0\times10^{-5}}$ | $\mathbf{1.0\times10^{-5}}$ | $\mathbf{3.0\times10^{-2}}$ | $\mathbf{4.0\times10^{-1}}$ | | $\mathbf{7.5\times10^{-5}}$ | $\mathbf{4.7\times10^{-4}}$ | $\mathbf{6.0\times10^{-4}}$ | $\mathbf{3.3\times10^{-4}}$ | **OK** |
| 11 | $1.0\times10^{-5}$ | $1.0\times10^{-5}$ | $5.0\times10^{-2}$ | $5.0\times10^{-2}$ | | $1.2\times10^{-4}$ | $2.9\times10^{-4}$ | $3.0\times10^{-4}$ | $4.0\times10^{-4}$ | |
| 12 | $1.0\times10^{-5}$ | $1.0\times10^{-5}$ | $5.0\times10^{-2}$ | $1.0\times10^{-1}$ | | $1.2\times10^{-4}$ | $3.3\times10^{-4}$ | $3.7\times10^{-4}$ | $4.5\times10^{-4}$ | |
| **13** | $\mathbf{1.0\times10^{-5}}$ | $\mathbf{1.0\times10^{-5}}$ | $\mathbf{5.0\times10^{-2}}$ | $\mathbf{3.0\times10^{-1}}$ | | $\mathbf{1.2\times10^{-4}}$ | $\mathbf{5.4\times10^{-4}}$ | $\mathbf{6.0\times10^{-4}}$ | $\mathbf{5.4\times10^{-4}}$ | **OK (wet Venus)** |
| **14** | $\mathbf{1.0\times10^{-5}}$ | $\mathbf{1.0\times10^{-5}}$ | $\mathbf{1.0\times10^{-1}}$ | $\mathbf{1.0\times10^{-1}}$ | | $\mathbf{2.3\times10^{-4}}$ | $\mathbf{5.7\times10^{-4}}$ | $\mathbf{5.8\times10^{-4}}$ | $\mathbf{7.9\times10^{-4}}$ | **OK (wet Venus)** |
| **15** | $\mathbf{1.0\times10^{-5}}$ | $\mathbf{1.0\times10^{-5}}$ | $\mathbf{1.0\times10^{-1}}$ | $\mathbf{3.0\times10^{-1}}$ | | $\mathbf{2.3\times10^{-4}}$ | $\mathbf{7.4\times10^{-4}}$ | $\mathbf{8.3\times10^{-4}}$ | $\mathbf{1.0\times10^{-3}}$ | **OK (wet Venus)** |
| 16 | $1.0\times10^{-5}$ | $1.0\times10^{-4}$ | $1.0\times10^{-3}$ | $5.0\times10^{-2}$ | $1.0\times10^{-1}$ | $2.4\times10^{-5}$ | $5.9\times10^{-5}$ | $7.9\times10^{-5}$ | $2.8\times10^{-5}$ | |
| 17 | $1.0\times10^{-5}$ | $1.0\times10^{-4}$ | $1.0\times10^{-3}$ | $1.0\times10^{-1}$ | | $2.4\times10^{-5}$ | $9.8\times10^{-5}$ | $1.2\times10^{-4}$ | $2.8\times10^{-5}$ | |
| 18 | $1.0\times10^{-5}$ | $1.0\times10^{-4}$ | $1.0\times10^{-3}$ | $3.0\times10^{-1}$ | | $2.4\times10^{-5}$ | $2.6\times10^{-4}$ | $3.2\times10^{-4}$ | $2.8\times10^{-5}$ | |
| 19 | $1.0\times10^{-5}$ | $1.0\times10^{-4}$ | $1.0\times10^{-2}$ | $5.0\times10^{-2}$ | | $3.9\times10^{-5}$ | $1.1\times10^{-4}$ | $1.3\times10^{-4}$ | $1.2\times10^{-4}$ | |
| 20 | $1.0\times10^{-5}$ | $1.0\times10^{-4}$ | $1.0\times10^{-2}$ | $1.0\times10^{-1}$ | | $3.9\times10^{-5}$ | $1.5\times10^{-4}$ | $1.8\times10^{-4}$ | $1.2\times10^{-4}$ | |
| 21 | $1.0\times10^{-5}$ | $1.0\times10^{-4}$ | $1.0\times10^{-2}$ | $3.0\times10^{-1}$ | | $3.9\times10^{-5}$ | $3.0\times10^{-4}$ | $3.9\times10^{-4}$ | $1.2\times10^{-4}$ | |
| 22 | $1.0\times10^{-5}$ | $1.0\times10^{-4}$ | $5.0\times10^{-2}$ | $5.0\times10^{-2}$ | | $1.3\times10^{-4}$ | $2.9\times10^{-4}$ | $3.0\times10^{-4}$ | $4.1\times10^{-4}$ | |
| 23 | $1.0\times10^{-5}$ | $1.0\times10^{-4}$ | $5.0\times10^{-2}$ | $1.0\times10^{-1}$ | | $1.3\times10^{-4}$ | $3.3\times10^{-4}$ | $3.7\times10^{-4}$ | $4.7\times10^{-4}$ | |
| **24** | $\mathbf{1.0\times10^{-5}}$ | $\mathbf{1.0\times10^{-4}}$ | $\mathbf{5.0\times10^{-2}}$ | $\mathbf{3.0\times10^{-1}}$ | | $\mathbf{1.3\times10^{-4}}$ | $\mathbf{5.4\times10^{-4}}$ | $\mathbf{6.1\times10^{-4}}$ | $\mathbf{5.5\times10^{-4}}$ | **OK (wet Venus)** |
| **25** | $\mathbf{1.0\times10^{-5}}$ | $\mathbf{1.0\times10^{-4}}$ | $\mathbf{1.0\times10^{-1}}$ | $\mathbf{1.0\times10^{-1}}$ | | $\mathbf{2.3\times10^{-4}}$ | $\mathbf{5.7\times10^{-4}}$ | $\mathbf{5.9\times10^{-4}}$ | $\mathbf{8.0\times10^{-4}}$ | **OK (wet Venus)** |
| **26** | $\mathbf{1.0\times10^{-5}}$ | $\mathbf{1.0\times10^{-4}}$ | $\mathbf{1.0\times10^{-1}}$ | $\mathbf{3.0\times10^{-1}}$ | | $\mathbf{2.3\times10^{-4}}$ | $\mathbf{7.5\times10^{-4}}$ | $\mathbf{8.3\times10^{-4}}$ | $\mathbf{1.0\times10^{-3}}$ | **OK (wet Venus)** |
| **27** | $\mathbf{1.0\times10^{-5}}$ | $\mathbf{5.0\times10^{-3}}$ | $\mathbf{4.3\times10^{-2}}$ | $\mathbf{6.5\times10^{-2}}$ | $\mathbf{1.2\times10^{-1}}$ | $\mathbf{6.4\times10^{-4}}$ | $\mathbf{5.2\times10^{-4}}$ | $\mathbf{5.9\times10^{-4}}$ | $\mathbf{8.0\times10^{-4}}$ | **OK (wet Venus)** |
| **28** | $\mathbf{1.0\times10^{-5}}$ | $\mathbf{1.0\times10^{-2}}$ | $\mathbf{7.5\times10^{-2}}$ | $\mathbf{1.1\times10^{-1}}$ | $\mathbf{1.9\times10^{-1}}$ | $\mathbf{1.2\times10^{-3}}$ | $\mathbf{9.6\times10^{-4}}$ | $\mathbf{1.1\times10^{-3}}$ | $\mathbf{1.5\times10^{-3}}$ | **OK (wet Venus)** |
| 29 | $1.0\times10^{-4}$ | $1.0\times10^{-4}$ | $1.0\times10^{-3}$ | $5.0\times10^{-2}$ | | $1.0\times10^{-4}$ | $1.4\times10^{-4}$ | $1.6\times10^{-4}$ | $1.1\times10^{-4}$ | |
| 30 | $1.0\times10^{-4}$ | $1.0\times10^{-4}$ | $1.0\times10^{-3}$ | $1.0\times10^{-1}$ | | $1.0\times10^{-4}$ | $1.8\times10^{-4}$ | $2.0\times10^{-4}$ | $1.1\times10^{-4}$ | |
| 31 | $1.0\times10^{-4}$ | $1.0\times10^{-4}$ | $1.0\times10^{-3}$ | $3.0\times10^{-1}$ | | $1.0\times10^{-4}$ | $3.4\times10^{-4}$ | $4.0\times10^{-4}$ | $1.1\times10^{-4}$ | |
| 32 | $1.0\times10^{-4}$ | $1.0\times10^{-4}$ | $1.0\times10^{-2}$ | $5.0\times10^{-2}$ | | $1.2\times10^{-4}$ | $2.0\times10^{-4}$ | $2.1\times10^{-4}$ | $2.0\times10^{-4}$ | |
| 33 | $1.0\times10^{-4}$ | $1.0\times10^{-4}$ | $1.0\times10^{-2}$ | $1.0\times10^{-1}$ | | $1.2\times10^{-4}$ | $2.3\times10^{-4}$ | $2.6\times10^{-4}$ | $2.0\times10^{-4}$ | |
| 34 | $1.0\times10^{-4}$ | $1.0\times10^{-4}$ | $1.0\times10^{-2}$ | $3.0\times10^{-1}$ | | $1.2\times10^{-4}$ | $3.8\times10^{-4}$ | $4.8\times10^{-4}$ | $2.0\times10^{-4}$ | |
| 35 | $1.0\times10^{-4}$ | $1.0\times10^{-4}$ | $3.0\times10^{-2}$ | $1.0\times10^{-1}$ | | $1.6\times10^{-4}$ | $3.3\times10^{-4}$ | $3.6\times10^{-4}$ | $4.0\times10^{-4}$ | |
| 36 | $1.0\times10^{-4}$ | $1.0\times10^{-4}$ | $3.0\times10^{-2}$ | $2.0\times10^{-1}$ | | $1.6\times10^{-4}$ | $4.3\times10^{-4}$ | $4.8\times10^{-4}$ | $4.2\times10^{-4}$ | |
| **37** | $\mathbf{1.0\times10^{-4}}$ | $\mathbf{1.0\times10^{-4}}$ | $\mathbf{3.0\times10^{-2}}$ | $\mathbf{3.0\times10^{-1}}$ | | $\mathbf{1.6\times10^{-4}}$ | $\mathbf{4.9\times10^{-4}}$ | $\mathbf{5.9\times10^{-4}}$ | $\mathbf{4.2\times10^{-4}}$ | **OK** |
| **38** | $\mathbf{1.0\times10^{-4}}$ | $\mathbf{1.0\times10^{-4}}$ | $\mathbf{3.0\times10^{-2}}$ | $\mathbf{4.0\times10^{-1}}$ | | $\mathbf{1.6\times10^{-4}}$ | $\mathbf{5.6\times10^{-4}}$ | $\mathbf{6.9\times10^{-4}}$ | $\mathbf{4.2\times10^{-4}}$ | **OK (wet Venus)** |
| 39 | $1.0\times10^{-4}$ | $1.0\times10^{-4}$ | $5.0\times10^{-2}$ | $5.0\times10^{-2}$ | | $2.1\times10^{-4}$ | $3.8\times10^{-4}$ | $3.9\times10^{-4}$ | $4.9\times10^{-4}$ | |
| 40 | $1.0\times10^{-4}$ | $1.0\times10^{-4}$ | $5.0\times10^{-2}$ | $1.0\times10^{-1}$ | | $2.1\times10^{-4}$ | $4.2\times10^{-4}$ | $4.6\times10^{-4}$ | $5.4\times10^{-4}$ | |
| **41** | $\mathbf{1.0\times10^{-4}}$ | $\mathbf{1.0\times10^{-4}}$ | $\mathbf{5.0\times10^{-2}}$ | $\mathbf{3.0\times10^{-1}}$ | | $\mathbf{2.1\times10^{-4}}$ | $\mathbf{6.2\times10^{-4}}$ | $\mathbf{6.9\times10^{-4}}$ | $\mathbf{6.3\times10^{-4}}$ | **OK (wet Venus)** |
| **42** | $\mathbf{1.0\times10^{-4}}$ | $\mathbf{1.0\times10^{-4}}$ | $\mathbf{1.0\times10^{-1}}$ | $\mathbf{1.0\times10^{-1}}$ | | $\mathbf{3.2\times10^{-4}}$ | $\mathbf{6.6\times10^{-4}}$ | $\mathbf{6.7\times10^{-4}}$ | $\mathbf{8.8\times10^{-4}}$ | **OK (wet Venus)** |
| **43** | $\mathbf{1.0\times10^{-4}}$ | $\mathbf{1.0\times10^{-4}}$ | $\mathbf{1.0\times10^{-1}}$ | $\mathbf{3.0\times10^{-1}}$ | | $\mathbf{3.2\times10^{-4}}$ | $\mathbf{8.3\times10^{-4}}$ | $\mathbf{9.2\times10^{-4}}$ | $\mathbf{1.1\times10^{-3}}$ | **OK (wet Venus)** |

| | | | | | | | | | |
|---|---|---|---|---|---|---|---|---|---|
| 44 | $1.0\times10^{-4}$ | $1.0\times10^{-3}$ | $1.0\times10^{-3}$ | $5.0\times10^{-2}$ | $1.9\times10^{-4}$ | $1.9\times10^{-4}$ | $2.2\times10^{-4}$ | $2.1\times10^{-4}$ | |
| 45 | $1.0\times10^{-4}$ | $1.0\times10^{-3}$ | $1.0\times10^{-3}$ | $1.0\times10^{-1}$ | $1.9\times10^{-4}$ | $2.3\times10^{-4}$ | $2.7\times10^{-4}$ | $2.1\times10^{-4}$ | |
| 46 | $1.0\times10^{-4}$ | $1.0\times10^{-3}$ | $1.0\times10^{-3}$ | $3.0\times10^{-1}$ | $1.9\times10^{-4}$ | $3.9\times10^{-4}$ | $4.9\times10^{-4}$ | $2.1\times10^{-4}$ | |
| 47 | $1.0\times10^{-4}$ | $1.0\times10^{-3}$ | $1.0\times10^{-2}$ | $5.0\times10^{-2}$ | $2.4\times10^{-4}$ | $2.4\times10^{-4}$ | $2.7\times10^{-4}$ | $2.8\times10^{-4}$ | |
| 48 | $1.0\times10^{-4}$ | $1.0\times10^{-3}$ | $1.0\times10^{-2}$ | $1.0\times10^{-1}$ | $2.4\times10^{-4}$ | $2.7\times10^{-4}$ | $3.3\times10^{-4}$ | $2.8\times10^{-4}$ | |
| **49** | **$1.0\times10^{-4}$** | **$1.0\times10^{-3}$** | **$1.0\times10^{-2}$** | **$3.0\times10^{-1}$** | **$2.4\times10^{-4}$** | **$4.2\times10^{-4}$** | **$5.4\times10^{-4}$** | **$2.8\times10^{-4}$** | **OK** |
| 50 | $1.0\times10^{-4}$ | $1.0\times10^{-3}$ | $3.0\times10^{-2}$ | $1.0\times10^{-1}$ | $2.7\times10^{-4}$ | $3.8\times10^{-4}$ | $4.2\times10^{-4}$ | $4.7\times10^{-4}$ | |
| **51** | **$1.0\times10^{-4}$** | **$1.0\times10^{-3}$** | **$3.0\times10^{-2}$** | **$2.0\times10^{-1}$** | **$2.7\times10^{-4}$** | **$4.8\times10^{-4}$** | **$5.5\times10^{-4}$** | **$5.0\times10^{-4}$** | **OK** |
| **52** | **$1.0\times10^{-4}$** | **$1.0\times10^{-3}$** | **$3.0\times10^{-2}$** | **$3.0\times10^{-1}$** | **$2.7\times10^{-4}$** | **$5.3\times10^{-4}$** | **$6.6\times10^{-4}$** | **$5.0\times10^{-4}$** | **OK (wet Venus)** |
| **53** | **$1.0\times10^{-4}$** | **$1.0\times10^{-3}$** | **$3.0\times10^{-2}$** | **$4.0\times10^{-1}$** | **$2.7\times10^{-4}$** | **$6.1\times10^{-4}$** | **$7.7\times10^{-4}$** | **$5.0\times10^{-4}$** | **OK (wet Venus)** |
| 54 | $1.0\times10^{-4}$ | $1.0\times10^{-3}$ | $5.0\times10^{-2}$ | $5.0\times10^{-2}$ | $2.8\times10^{-4}$ | $4.3\times10^{-4}$ | $4.3\times10^{-4}$ | $5.5\times10^{-4}$ | |
| **55** | **$1.0\times10^{-4}$** | **$1.0\times10^{-3}$** | **$5.0\times10^{-2}$** | **$1.0\times10^{-1}$** | **$2.8\times10^{-4}$** | **$4.6\times10^{-4}$** | **$5.0\times10^{-4}$** | **$6.6\times10^{-4}$** | **OK** |
| **56** | **$1.0\times10^{-4}$** | **$1.0\times10^{-3}$** | **$5.0\times10^{-2}$** | **$3.0\times10^{-1}$** | **$2.8\times10^{-4}$** | **$6.7\times10^{-4}$** | **$7.7\times10^{-4}$** | **$7.0\times10^{-4}$** | **OK (wet Venus)** |
| **57** | **$1.0\times10^{-4}$** | **$1.0\times10^{-3}$** | **$1.0\times10^{-1}$** | **$1.0\times10^{-1}$** | **$3.9\times10^{-4}$** | **$7.1\times10^{-4}$** | **$7.1\times10^{-4}$** | **$9.4\times10^{-4}$** | **OK (wet Venus)** |
| **58** | **$1.0\times10^{-4}$** | **$1.0\times10^{-3}$** | **$1.0\times10^{-1}$** | **$3.0\times10^{-1}$** | **$3.9\times10^{-4}$** | **$8.8\times10^{-4}$** | **$9.7\times10^{-4}$** | **$1.2\times10^{-3}$** | **OK (wet Venus)** |
| **59** | **$1.0\times10^{-4}$** | **$5.0\times10^{-3}$** | **$5.0\times10^{-3}$** | **$1.0\times10^{-1}$** | **$5.8\times10^{-4}$** | **$4.5\times10^{-4}$** | **$5.7\times10^{-4}$** | **$6.4\times10^{-4}$** | **OK** |
| **60** | **$1.0\times10^{-4}$** | **$5.0\times10^{-3}$** | **$5.0\times10^{-3}$** | **$2.0\times10^{-1}$** | **$5.8\times10^{-4}$** | **$5.4\times10^{-4}$** | **$6.8\times10^{-4}$** | **$6.9\times10^{-4}$** | **OK (wet Venus)** |
| **61** | **$1.0\times10^{-4}$** | **$5.0\times10^{-3}$** | **$5.0\times10^{-3}$** | **$3.0\times10^{-1}$** | **$5.8\times10^{-4}$** | **$6.2\times10^{-4}$** | **$8.1\times10^{-4}$** | **$6.9\times10^{-4}$** | **OK (wet Venus)** |
| **62** | **$1.0\times10^{-4}$** | **$5.0\times10^{-3}$** | **$5.0\times10^{-3}$** | **$4.0\times10^{-1}$** | **$5.8\times10^{-4}$** | **$7.0\times10^{-4}$** | **$9.1\times10^{-4}$** | **$6.9\times10^{-4}$** | **OK (wet Venus)** |
| **63** | **$1.0\times10^{-4}$** | **$5.0\times10^{-3}$** | **$1.0\times10^{-2}$** | **$1.0\times10^{-1}$** | **$5.9\times10^{-4}$** | **$4.7\times10^{-4}$** | **$5.9\times10^{-4}$** | **$7.0\times10^{-4}$** | **OK** |
| **64** | **$1.0\times10^{-4}$** | **$5.0\times10^{-3}$** | **$1.0\times10^{-2}$** | **$2.0\times10^{-1}$** | **$5.9\times10^{-4}$** | **$5.6\times10^{-4}$** | **$7.1\times10^{-4}$** | **$7.2\times10^{-4}$** | **OK (wet Venus)** |
| **65** | **$1.0\times10^{-4}$** | **$5.0\times10^{-3}$** | **$1.0\times10^{-2}$** | **$3.0\times10^{-1}$** | **$5.9\times10^{-4}$** | **$6.4\times10^{-4}$** | **$8.2\times10^{-4}$** | **$7.2\times10^{-4}$** | **OK (wet Venus)** |
| **66** | **$1.0\times10^{-4}$** | **$5.0\times10^{-3}$** | **$2.0\times10^{-2}$** | **$5.0\times10^{-2}$** | **$6.3\times10^{-4}$** | **$4.8\times10^{-4}$** | **$5.5\times10^{-4}$** | **$7.4\times10^{-4}$** | **OK** |
| 67 | $1.0\times10^{-4}$ | $5.0\times10^{-3}$ | $2.0\times10^{-2}$ | $1.0\times10^{-1}$ | $6.9\times10^{-4}$ | $5.2\times10^{-4}$ | $6.2\times10^{-4}$ | $7.6\times10^{-4}$ | OK (wet Venus) |
| 68 | $1.0\times10^{-4}$ | $5.0\times10^{-3}$ | $2.0\times10^{-2}$ | $2.0\times10^{-1}$ | $6.9\times10^{-4}$ | $6.1\times10^{-4}$ | $7.7\times10^{-4}$ | $7.8\times10^{-4}$ | OK (wet Venus) |
| 69 | $1.0\times10^{-4}$ | $5.0\times10^{-3}$ | $2.0\times10^{-2}$ | $3.0\times10^{-1}$ | $6.9\times10^{-4}$ | $6.9\times10^{-4}$ | $8.8\times10^{-4}$ | $7.8\times10^{-4}$ | OK (wet Venus) |
| 70 | $1.0\times10^{-4}$ | $5.0\times10^{-3}$ | $3.0\times10^{-2}$ | $5.0\times10^{-2}$ | $6.9\times10^{-4}$ | $5.3\times10^{-4}$ | $6.0\times10^{-4}$ | $7.8\times10^{-4}$ | OK (wet Venus) |
| 71 | $1.0\times10^{-4}$ | $5.0\times10^{-3}$ | $3.0\times10^{-2}$ | $1.0\times10^{-1}$ | $7.0\times10^{-4}$ | $5.7\times10^{-4}$ | $6.5\times10^{-4}$ | $8.0\times10^{-4}$ | OK (wet Venus) |
| 72 | $1.0\times10^{-4}$ | $5.0\times10^{-3}$ | $3.0\times10^{-2}$ | $2.0\times10^{-1}$ | $7.0\times10^{-4}$ | $6.7\times10^{-4}$ | $8.1\times10^{-4}$ | $8.5\times10^{-4}$ | OK (wet Venus) |
| 73 | $1.0\times10^{-4}$ | $5.0\times10^{-3}$ | $3.0\times10^{-2}$ | $3.0\times10^{-1}$ | $7.0\times10^{-4}$ | $7.5\times10^{-4}$ | $9.5\times10^{-4}$ | $8.5\times10^{-4}$ | OK (wet Venus) |
| 74 | $1.0\times10^{-4}$ | $5.0\times10^{-3}$ | $3.0\times10^{-2}$ | $4.0\times10^{-1}$ | $7.0\times10^{-4}$ | $8.3\times10^{-4}$ | $1.1\times10^{-3}$ | $8.5\times10^{-4}$ | OK (wet Venus) |
| 75 | $1.0\times10^{-4}$ | $5.0\times10^{-3}$ | $5.0\times10^{-2}$ | $1.0\times10^{-1}$ | $8.1\times10^{-4}$ | $6.6\times10^{-4}$ | $7.5\times10^{-4}$ | $9.6\times10^{-4}$ | OK (wet Venus) |
| 76 | $1.0\times10^{-4}$ | $5.0\times10^{-3}$ | $5.0\times10^{-2}$ | $2.0\times10^{-1}$ | $8.1\times10^{-4}$ | $7.6\times10^{-4}$ | $8.8\times10^{-4}$ | $1.0\times10^{-3}$ | OK (wet Venus) |
| 77 | $1.0\times10^{-4}$ | $5.0\times10^{-3}$ | $5.0\times10^{-2}$ | $3.0\times10^{-1}$ | $8.1\times10^{-4}$ | $8.6\times10^{-4}$ | $1.0\times10^{-3}$ | $1.0\times10^{-3}$ | OK (wet Venus) |
| 78 | $1.0\times10^{-4}$ | $5.0\times10^{-3}$ | $1.0\times10^{-1}$ | $1.0\times10^{-1}$ | $9.4\times10^{-4}$ | $9.1\times10^{-4}$ | $9.5\times10^{-4}$ | $1.3\times10^{-3}$ | OK (wet Venus) |
| 79 | $1.0\times10^{-4}$ | $5.0\times10^{-3}$ | $1.0\times10^{-1}$ | $2.0\times10^{-1}$ | $9.4\times10^{-4}$ | $9.8\times10^{-4}$ | $1.1\times10^{-3}$ | $1.4\times10^{-3}$ | OK (wet Venus) |
| 80 | $1.0\times10^{-4}$ | $5.0\times10^{-3}$ | $1.0\times10^{-1}$ | $3.0\times10^{-1}$ | $9.4\times10^{-4}$ | $1.1\times10^{-3}$ | $1.2\times10^{-3}$ | $1.5\times10^{-3}$ | OK (wet Venus) |

**Notes.** Successful systems are highlighted in bold. Model 1 was adopted in ref.[4], the most widely used WMF model in the literature. Model 2 is based on model 1 and was often used in the instability models of ref.[9]. Model 16 was introduced in the review of ref.[3] and used in the models of ref.[19]. The four columns on the right side show the median WMFs acquired by each of the planet analogues identified in all systems in the standard disk. If the median WMFs of Venus, Earth and Mars simultaneously satisfied the respective observational constraints ($0.1–5 \times 10^{-4}$, $5–25 \times 10^{-4}$ and $0.5–20 \times 10^{-4}$, respectively), the WMF model was deemed successful. Here, 'wet Venus' refers to the assumption that early Venus was wet by setting its maximum WMF constraint to $5 \times 10^{-3}$. See Methods and Supplementary Information for more details.

Table S3. Summary of key variables for the 47 systems simultaneously containing analogues of Mercury, Venus, Earth and Mars

| System # | Disk | AMD | RMC | tLGI (Myr) | Lvf (%) | tMars (Myr) | C1 | C2 | C3 | C4 | C5 | Result |
|---|---|---|---|---|---|---|---|---|---|---|---|---|
| 1 | D | 0.0094 | 53.4 | 17 | 4.2 | 178.2 | X | O | Δ | Δ | X | 2 |
| 2 | Ib | 0.0008 | 91.9 | 18 | 11.1 | 138.9 | O | O | Δ | X | X | 2 |
| 3 | Ia | 0.0024 | 75.9 | 8 | 21.0 | 373.2 | O | O | X | X | X | 3 |
| **4** | **Ia** | **0.0008** | **75.6** | **146** | **0.7** | **33.9** | **O** | **O** | **O** | **O** | **Δ** | **OK** |
| 5 | Ia | 0.0007 | 80.4 | 3 | 29.5 | 60.4 | O | O | X | X | X | 3 |
| 6 | Ie | 0.0029 | 56.7 | 10 | 12.3 | 22.9 | O | O | X | X | Δ | 2 |
| **7** | **Ib** | **0.0059** | **71.0** | **206** | **0.2** | **9.7** | **Δ** | **O** | **O** | **O** | **O** | **OK** |
| **8** | **Ib** | **0.0033** | **53.3** | **365** | **0.0** | **11.8** | **O** | **O** | **X** | **O** | **O** | **marginally ok** |
| 9 | Id | 0.003 | 52.2 | 2 | 41.2 | 22.6 | O | O | X | X | Δ | 2 |
| **10** | **Ib** | **0.0026** | **53.8** | **53** | **2.5** | **24.0** | **O** | **O** | **O** | **Δ** | **Δ** | **OK** |
| **11** | **Id** | **0.0009** | **49.2** | **36** | **5.7** | **31.0** | **O** | **O** | **O** | **X** | **Δ** | **marginally ok** |
| **12** | **Ic** | **0.0018** | **70.1** | **151** | **0.7** | **282.3** | **O** | **O** | **O** | **O** | **X** | **marginally ok** |
| 13 | A | 0.006 | 68.4 | 12 | 12.6 | 20.2 | Δ | O | X | X | Δ | 2 |
| **14** | **B** | **0.0028** | **102.3** | **18** | **4.0** | **6.7** | **O** | **O** | **Δ** | **Δ** | **O** | **OK** |
| **15** | **C** | **0.0032** | **63.9** | **49** | **0.9** | **7.7** | **O** | **O** | **O** | **O** | **O** | **OK** |
| **16** | **Id** | **0.002** | **50.3** | **166** | **0.4** | **15.0** | **O** | **O** | **O** | **O** | **O** | **OK** |
| **17** | **Id** | **0.0092** | **43.0** | **81** | **0.8** | **27.7** | **X** | **Δ** | **O** | **O** | **Δ** | **marginally ok** |
| **18** | **Ie** | **0.0059** | **48.8** | **22** | **5.2** | **21.7** | **Δ** | **O** | **Δ** | **Δ** | **Δ** | **OK** |
| **19** | **Ib** | **0.0033** | **50.2** | **93** | **0.8** | **25.3** | **O** | **O** | **O** | **O** | **Δ** | **OK** |
| **20** | **Ia** | **0.0054** | **50.4** | **100** | **0.9** | **154.2** | **Δ** | **O** | **O** | **O** | **X** | **marginally ok** |
| **21** | **D** | **0.0021** | **64.7** | **35** | **1.3** | **13.2** | **O** | **O** | **O** | **Δ** | **O** | **OK** |
| **22** | **Ia** | **0.0008** | **77.0** | **45** | **4.4** | **33.9** | **O** | **O** | **O** | **Δ** | **Δ** | **OK** |
| **23** | **B** | **0.0066** | **70.7** | **48** | **1.6** | **201.2** | **Δ** | **O** | **O** | **Δ** | **X** | **marginally ok** |
| 24 | Ic | 0.0016 | 55.9 | 25 | 8.1 | 57.7 | O | O | O | X | X | 2 |

| | | | | | | | | | | | |
|---|---|---|---|---|---|---|---|---|---|---|---|
| 25 | Ia | 0.015 | 43.4 | 396 | 0.0 | 46.2 | X | Δ | X | O | X | 3 |
| 26 | Ic | 0.003 | 38.5 | 23 | 10.7 | 33.9 | O | X | Δ | X | Δ | 2 |
| **27** | **Id** | **0.0011** | **40.4** | **54** | **3.7** | **27.2** | **O** | **Δ** | **O** | **Δ** | **Δ** | **OK** |
| 28 | Ia | 0.0009 | 69.5 | 8 | 18.2 | 54.6 | O | O | X | X | X | 3 |
| **29** | **D** | **0.0091** | **52.4** | **51** | **1.0** | **13.5** | **X** | **O** | **O** | **O** | **O** | **marginally ok** |
| **30** | **Ib** | **0.0015** | **52.5** | **138** | **0.2** | **53.2** | **O** | **O** | **O** | **O** | **X** | **marginally ok** |
| **31** | **Ib** | **0.003** | **52.8** | **166** | **0.3** | **12.2** | **O** | **O** | **O** | **O** | **O** | **OK** |
| **32** | **Ib** | **0.005** | **40.0** | **80** | **0.9** | **9.1** | **Δ** | **Δ** | **O** | **O** | **O** | **OK** |
| 33 | C | 0.0011 | 67.0 | 8 | 12.2 | 4.4 | O | O | X | X | O | 2 |
| **34** | **A** | **0.0021** | **61.1** | **43** | **1.3** | **18.4** | **O** | **O** | **O** | **Δ** | **O** | **OK** |
| **35** | **B** | **0.0011** | **56.9** | **30** | **3.2** | **8.1** | **O** | **O** | **O** | **Δ** | **O** | **OK** |
| **36** | **C** | **0.0012** | **67.8** | **56** | **1.0** | **90.9** | **O** | **O** | **O** | **O** | **X** | **marginally ok** |
| **37** | **D** | **0.0014** | **63.5** | **20** | **3.1** | **6.9** | **O** | **O** | **Δ** | **Δ** | **O** | **OK** |
| 38 | Ia | 0.0024 | 63.1 | 10 | 17.6 | 136.1 | O | O | X | X | X | 3 |
| **39** | **Ie** | **0.0024** | **50.7** | **79** | **0.7** | **11.8** | **O** | **O** | **O** | **O** | **O** | **OK** |
| 40 | Id | 0.0012 | 46.4 | 10 | 18.5 | 29.4 | O | O | X | X | Δ | 2 |
| **41** | **C** | **0.0026** | **110.2** | **204** | **0.0** | **161.1** | **O** | **O** | **O** | **O** | **X** | **marginally ok** |
| 42 | Ic | 0.0014 | 51.6 | 13 | 13.5 | 35.3 | O | O | X | X | X | 3 |
| **43** | **Ie** | **0.0031** | **49.2** | **103** | **0.1** | **23.3** | **O** | **O** | **O** | **O** | **Δ** | **OK** |
| 44 | Ib | 0.0008 | 51.0 | 33 | 7.0 | 36.6 | O | O | O | X | X | 2 |
| 45 | C | 0.0032 | 69.1 | 6 | 16.2 | 261.9 | O | O | X | X | X | 3 |
| **46** | **Ie** | **0.0023** | **51.6** | **43** | **2.7** | **19.4** | **O** | **O** | **O** | **Δ** | **Δ** | **OK** |
| **47** | **Ic** | **0.0009** | **55.4** | **64** | **2.3** | **24.3** | **O** | **O** | **O** | **Δ** | **Δ** | **OK** |
| **Solar system** | - | 0.0018 | 89.7 | 25-245 | <1 | <15-23 | - | - | - | - | - | |

**Notes.** Successful systems are highlighted in bold. AMD, angular momentum deficit; RMC, radial mass concentration; tLGI, the time of the last giant impact of the Earth analogue (defined here by the collision of an object 5% as massive as the target body); Lvf, late veneer mass fraction of the representative Earth analogue; tMars, formation time of Mars analogues. Except for the number of analogue systems, all other quantities are represented by medians in each disk model. 'O' and 'Δ' indicate that a system fully and marginally satisfied a specific constraint among C1–C5, respectively, while 'X' indicates that it did not. The constraints were defined as C1: AMD (0–0.0036), C2: RMC (44.9–179.4), C3: tLGI for the Earth analogue (25–245 Myr), C4: LVf for the Earth analogue (< 1%), C5: Mars analogue formation time (< 15–23 Myr). The water-mass fractions of all Earth analogues satisfied the observed constraints in several reasonable WMF models (Table S2). Most Mars analogues that failed to meet tMars suffered one giant impact by 30–40 Myr, so it was not the result of a continuous accretion history. See Supplementary Information for more details.

**Table S4.** Representative analogues of Mercury, Venus, Earth and Mars formed in all our terrestrial planet analogue systems

|  | N | a (au) | e | i (°) | m (M⊕) | TF 80% [90%] (Myr) |
|---|---|---|---|---|---|---|
| *Mercury* | | | | | | |
| A | 7 | 0.510 | 0.103 | 5.616 | 0.144 | |
| B | 9 | 0.492 | 0.107 | 4.712 | 0.112 | |
| C | 9 | 0.475 | 0.115 | 3.947 | 0.107 | |
| D | 9 | 0.492 | 0.133 | 6.947 | 0.188 | |
| E | 4 | 0.467 | 0.081 | 2.633 | 0.259 | |
| Ia | 23 | 0.465 | 0.061 | 3.220 | 0.109 | |
| Ib | 18 | 0.452 | 0.073 | 3.182 | 0.154 | |
| Ic | 12 | 0.420 | 0.085 | 3.720 | 0.112 | |
| Id | 15 | 0.448 | 0.072 | 3.715 | 0.203 | |
| Ie | 16 | 0.400 | 0.106 | 4.356 | 0.139 | |
| Standard (ABC) | 25 | 0.492 | 0.112 | 4.712 | 0.112 | |
| A–E | 38 | 0.488 | 0.110 | 4.709 | 0.139 | |
| Ix | 84 | 0.442 | 0.074 | 3.545 | 0.158 | |
| 4-P systems | 47 | 0.467 | 0.078 | 3.947 | 0.163 | |
| *Venus* | | | | | | |
| A | 21 | 0.649 | 0.040 | 2.222 | 0.925 | |
| B | 30 | 0.656 | 0.044 | 2.320 | 0.926 | |
| C | 36 | 0.653 | 0.041 | 2.146 | 0.906 | |
| D | 13 | 0.697 | 0.048 | 2.165 | 1.157 | |
| E | 10 | 0.645 | 0.046 | 2.115 | 0.989 | |
| Ia | 29 | 0.653 | 0.034 | 1.991 | 0.769 | |
| Ib | 25 | 0.675 | 0.039 | 2.156 | 0.819 | |
| Ic | 18 | 0.690 | 0.026 | 1.915 | 1.000 | |
| Id | 21 | 0.667 | 0.035 | 2.114 | 0.822 | |
| Ie | 18 | 0.663 | 0.040 | 2.347 | 0.993 | |
| Standard (ABC) | 87 | 0.654 | 0.042 | 2.222 | 0.925 | |
| A–E | 110 | 0.657 | 0.043 | 2.207 | 0.952 | |
| Ix | 111 | 0.669 | 0.035 | 2.097 | 0.822 | |
| 4-P systems | 47 | 0.680 | 0.031 | 2.154 | 0.998 | |
| *Earth* | | | | | | |
| A | 21 | 1.067 | 0.050 | 2.027 | 1.196 | 17.7 [24.5] |
| B | 30 | 1.064 | 0.035 | 2.088 | 1.035 | 15.7 [22.4] |
| C | 36 | 1.045 | 0.043 | 2.124 | 0.873 | 18.6 [32.7] |
| D | 13 | 1.137 | 0.052 | 2.568 | 1.030 | 20.2 [28.0] |
| E | 10 | 1.046 | 0.050 | 1.931 | 0.890 | 83.0 [91.2] |
| Ia | 29 | 0.948 | 0.025 | 2.033 | 0.927 | 32.0 [52.4] |
| Ib | 25 | 0.987 | 0.029 | 2.037 | 0.892 | 18.4 [30.5] |
| Ic | 18 | 1.078 | 0.028 | 2.063 | 0.777 | 22.0 [33.4] |
| Id | 21 | 0.996 | 0.032 | 2.121 | 0.892 | 21.5 [35.6] |
| Ie | 18 | 1.010 | 0.035 | 2.334 | 0.867 | 19.3 [29.3] |
| Standard (ABC) | 87 | 1.063 | 0.043 | 2.109 | 1.033 | 16.5 [28.1] |
| A–E | 110 | 1.065 | 0.044 | 2.130 | 0.999 | 18.1 [29.8] |
| Ix | 111 | 0.996 | 0.030 | 2.094 | 0.858 | 20.0 [35.6] |
| 4-P systems | 47 | 1.048 | 0.027 | 2.044 | 0.898 | 17.7 [33.2] |
| *Mars* | | | | | | |
| A | 16 | 1.497 | 0.095 | 6.032 | 0.147 | 14.2 [20.1] |
| B | 24 | 1.560 | 0.099 | 7.491 | 0.130 | 16.9 [20.0] |
| C | 32 | 1.591 | 0.073 | 5.632 | 0.126 | 17.3 [19.2] |
| D | 8 | 1.683 | 0.099 | 5.789 | 0.174 | 10.0 [13.4] |
| E | 6 | 1.622 | 0.097 | 4.635 | 0.119 | 1.0 [9.1] |
| Ia | 14 | 1.479 | 0.057 | 3.367 | 0.212 | 36.1 [41.8] |
| Ib | 16 | 1.596 | 0.073 | 5.989 | 0.228 | 24.0 [24.7] |
| Ic | 11 | 1.691 | 0.038 | 4.546 | 0.128 | 18.1 [33.6] |
| Id | 12 | 1.535 | 0.057 | 3.275 | 0.220 | 13.2 [27.5] |
| Ie | 7 | 1.637 | 0.099 | 5.146 | 0.106 | 6.2 [21.4] |
| Standard (ABC) | 72 | 1.545 | 0.080 | 6.201 | 0.133 | 15.5 [19.7] |
| A–E | 86 | 1.578 | 0.089 | 6.143 | 0.136 | 14.9 [17.9] |
| Ix | 60 | 1.565 | 0.066 | 4.319 | 0.203 | 18.3 [27.5] |
| 4-P systems | 47 | 1.556 | 0.058 | 4.534 | 0.150 | 18.4 [27.2] |
| **Mercury** | … | 0.387 | 0.215 | 6.784 | 0.055 | |
| **Venus** | … | 0.723 | 0.032 | 2.159 | 0.815 | |
| **Earth** | … | 1.000 | 0.028 | 1.988 | 1.000 | |
| **Mars** | … | 1.524 | 0.067 | 4.042 | 0.107 | |

**Notes.** N, number of identified representative planet analogues; $a$, the semimajor axis; $e$, eccentricity; $i$, inclination; $m$, planet mass; TF, the time the planet acquired 80% or 90% of its final mass. Standard disk (ABC), A–E and Ix show the results for disks combined (e.g., disk Ix refers to the results of disks Ia, Ib, Ic, Id and Ie combined). A–E and Ix represent disks with small and extended inner regions, respectively.

The four representative terrestrial-planet analogues are present in four-planet systems. Four-planet systems are taken altogether when determining the various quantities, regardless of the disk model from which these systems originated. All quantities (except N) represent median values. We averaged the orbital elements of the real planets over the last 100 Myr after integrating their orbits. See Methods and Supplementary Information for more details.

Table S5. Summary of key variables obtained for all of our obtained terrestrial-planet analogue systems

| Disk | N0 | n4 | AMD | RMC | tLGI (Myr) | Lvf (%) | tMars (Myr) | C1 (%) | C2 (%) | C3 (%) | C4 (%) | C5 (%) |
|---|---|---|---|---|---|---|---|---|---|---|---|---|
| A | 100 | 2 | 0.0031 | 61.5 | 26 | 3.3 | 20 | 62 | 95 | 52 | 10 | 33 |
| B | 100 | 3 | 0.0030 | 67.8 | 20 | 4.0 | 20 | 60 | 100 | 40 | 27 | 33 |
| C | 100 | 5 | 0.0030 | 63.4 | 45 | 2.1 | 19 | 61 | 94 | 67 | 44 | 28 |
| D | 50 | 4 | 0.0029 | 56.9 | 33 | 2.1 | 13 | 54 | 100 | 77 | 23 | 38 |
| E | 50 | 0 | 0.0033 | 44.7 | 91 | 0.5 | 9 | 50 | 60 | 70 | 70 | 60 |
| Ia | 50 | 8 | 0.0018 | 55.7 | 67 | 2.1 | 42 | 76 | 93 | 62 | 48 | 0 |
| Ib | 50 | 9 | 0.0023 | 53.3 | 47 | 3.0 | 25 | 76 | 88 | 52 | 44 | 20 |
| Ic | 50 | 5 | 0.0016 | 53.5 | 30 | 6.5 | 34 | 83 | 94 | 72 | 17 | 11 |
| Id | 50 | 6 | 0.0020 | 45.7 | 36 | 5.7 | 27 | 67 | 52 | 52 | 33 | 10 |
| Ie | 50 | 5 | 0.0025 | 50.5 | 32 | 5.1 | 21 | 83 | 89 | 56 | 39 | 17 |
| Standard (ABC) | 300 | 10 | 0.0030 | 64.3 | 30 | 2.9 | 20 | 61 | 97 | 54 | 30 | 31 |
| A–E | 400 | 14 | 0.0030 | 61.8 | 37 | 2.7 | 18 | 59 | 94 | 58 | 33 | 35 |
| Ix | 250 | 33 | 0.0020 | 52.2 | 44 | 4.0 | 27 | 77 | 84 | 59 | 38 | 11 |
| 4-P systems | 650 | 47 | 0.0024 | 55.4 | 43 | 2.7 | 27 | 79 | 89 | 60 | 38 | 30 |
| **Solar system** | - | - | **0.0018** | **89.7** | **25-245** | **<1** | **15-23** | - | - | - | - | - |

**Notes.** N0, the number of simulation runs performed for a given disk model; n4, the number of analogue systems containing Mercury–Venus–Earth–Mars representative analogues. Disk models and other variables are described in the captions of Tables S3 and S4. Except for the number of analogue systems, all other quantities are represented by medians found in each disk model. The last five columns represent the successful fractions of analogue systems that satisfied a particular constraint. See Methods and Supplementary Information for more details.